\documentclass[onecolumn,superscriptaddress,nofootinbib,notitlepage]{revtex4-1}
\pdfoutput=1

\usepackage{graphicx}
\usepackage{amsmath}
\usepackage{amssymb}
\usepackage{amsfonts}
\usepackage{dcolumn}
\usepackage{bm}
\usepackage[linktoc=none]{hyperref}
\usepackage[dvipsnames]{xcolor}
\usepackage[utf8]{inputenc}
\usepackage{subfigure}

\usepackage[T1]{fontenc}
\usepackage{pstricks}
\usepackage{color}
\usepackage{multirow}
\usepackage{slashed}

\usepackage{braket}
\usepackage{url}
\usepackage{relsize}
\usepackage{fullpage}
\usepackage{makecell}
\usepackage{blkarray}
\usepackage{fullpage}

\usepackage{feynmp}
\usepackage{feynmp-auto}
\hypersetup{colorlinks,linkcolor={blue},citecolor={teal},urlcolor={violet}}  

\newcommand{\GRAPPA}{%
Gravitation Astroparticle Physics Amsterdam (GRAPPA),\\
Institute for Theoretical Physics Amsterdam
and Delta Institute for Theoretical Physics,\\
University of Amsterdam, Science Park 904, 1098 XH Amsterdam, The Netherlands}

\newcommand{\SOTON}{Department of Physics and Astronomy, University of Southampton, SO17 1BJ Southampton, United Kingdom}

\begin{document}

\title{Minimal Seesaw extension for Neutrino Mass and Mixing, Leptogenesis and Dark Matter: FIMPzillas through the Right-Handed Neutrino Portal}

\author{Marco Chianese}
\email{m.chianese@uva.nl}
\affiliation{\GRAPPA}

\author{Bowen Fu}
\email{B.Fu@soton.ac.uk}
\affiliation{\SOTON}

\author{Stephen F. King}
\email{king@soton.ac.uk}
\affiliation{\SOTON}

\date{\today}

\begin{abstract}
We propose a minimal seesaw extension to simultaneously account for realistic neutrino mass and mixing, the baryon asymmetry of the Universe via leptogenesis and a viable dark matter relic density, in which two right-handed neutrinos are coupled to a dark Dirac fermion and complex scalar field, both charged under a global $U(1)_D$ symmetry. As a concrete example, we consider the Littlest Seesaw model which describes neutrino mass and mixing and accounts for leptogenesis, thereby fixing the neutrino Yukawa couplings and right-handed neutrino masses. By considering the freeze-in production mechanism of dark matter, we explore the parameter space of right-handed neutrino portal couplings and dark particle masses which give the correct dark matter relic abundance, focussing on the case of a superheavy Dirac fermion dark matter particle, with a mass around $10^{10}$ GeV. Such a FIMPzilla can provide a successful explanation of the dark matter relic abundance, with its production reliant on neutrino Yukawa couplings over much of the parameter space, depending on the assumed dark particle masses, and the reheat temperature.
\end{abstract}

\maketitle

\tableofcontents

\section{Introduction\label{sec:intro}}

The masses of neutrinos and their mixing, evidenced by the neutrino oscillation experiments \cite{2016NuPhB.908....1O}, represent the most convincing physics beyond the Standard Model. Although the origin of neutrino mass and mixing is unknown, one of the most widely studied candidate explanations is the well-known seesaw mechanism, proposed in the late 1970s \cite{Minkowski:1977sc,Yanagida:1979as,GellMann:1980vs,Schechter:1980gr,Mohapatra:1979ia,Mohapatra:1980yp}. However, the most general version of seesaw model has many free parameters. In general, there are not enough physical constraints to fix the parameters, and thus the model is hard to be tested even indirectly. A natural and effective solution to reduce the number of free parameters is to consider only two right-handed neutrinos (2RHN) with one texture zero \cite{King:1999mb, King:2002nf}, in which the lightest neutrino has zero mass. The number of free parameters could be further reduced by imposing two texture zeros in the Dirac neutrino mass matrix \cite{Frampton:2002qc}. However, such a two texture zero model is incompatible with the normal hierarchy of neutrino masses even though the consistency with cosmological leptogenesis is kept \cite{Fukugita:1986hr,Guo:2003cc, Ibarra:2003up, Mei:2003gn, Guo:2006qa, Antusch:2011nz,Harigaya:2012bw, Zhang:2015tea}, while the one texture zero model is compatible with the normal neutrino mass hierarchy.

The Littlest Seesaw (LS) model is based on the one texture zero 2RHN model with a constrained sequential dominance (CSD) form of Dirac neutrino mass matrix where $n=3$ \cite{King:2013iva, Bjorkeroth:2014vha, King:2015dvf,Bjorkeroth:2015ora,Bjorkeroth:2015tsa,King:2016yvg,Ballett:2016yod}. The number of independent Yukawa couplings is only two, which means the model is highly predictive. Recently, an attempt has been made trying to use currently constrained observables the low-energy neutrino data and leptogenesis to fit the high-energy LS parameters \cite{King:2018fqh}. The result turns out to be good, with the best fit $\Delta \chi^2$ being 1.51 for three degrees of freedom. However, such a model by itself does not address the problem of cosmological Dark Matter (DM).

There are many works trying to relate the DM problem to the neutrino mass and mixing problem \cite{Caldwell:1993kn,Mohapatra:2002ug,Krauss:2002px,Ma:2006km,Asaka:2005an,Boehm:2006mi,Kubo:2006yx,Ma:2006fn,Hambye:2006zn,Lattanzi:2007ux,Ma:2007gq,Allahverdi:2007wt,Gu:2007ug,Sahu:2008aw,Arina:2008bb,Aoki:2008av,Ma:2008cu,Gu:2008yj,Aoki:2009vf,Gu:2010yf,Hirsch:2010ru,Esteves:2010sh,Kanemura:2011vm,Lindner:2011it,JosseMichaux:2011ba,Schmidt:2012yg,Borah:2012qr,Farzan:2012sa,Chao:2012mx,Gustafsson:2012vj,Blennow:2013pya,Law:2013saa,Hernandez:2013dta,Restrepo:2013aga,Chakraborty:2013gea,Ahriche:2014cda,Kanemura:2014rpa,Huang:2014bva,Varzielas:2015joa,Sanchez-Vega:2015qva,Fraser:2015mhb,Adhikari:2015woo,Ahriche:2016rgf,Sierra:2016qfa,Lu:2016ucn,Batell:2016zod,Ho:2016aye,Escudero:2016ksa,Bonilla:2016diq,Borah:2016zbd,Biswas:2016yan,Hierro:2016nwm,Bhattacharya:2016qsg,Chakraborty:2017dfg,Bhattacharya:2017sml,Ho:2017fte,Ghosh:2017fmr,Nanda:2017bmi,Narendra:2017uxl,Bernal:2017xat,Borah:2018gjk,Batell:2017cmf,Pospelov:2007mp,Falkowski:2009yz,Falkowski:2011xh,Cherry:2014xra,Bertoni:2014mva,Allahverdi:2016fvl,Karam:2015jta,Bhattacharya:2018ljs}. A recent example is presented in \cite{Chianese:2018dsz}, which proposes a simple extension of the LS model. The model in \cite{Chianese:2018dsz} includes a scalar dark particle and a Dirac spinor dark matter, both of which are odd under an additional global $U(1)_{D}$ symmetry.\footnote{Actually the model in \cite{Chianese:2018dsz} suggested a dark $Z_2$ parity but in fact the restricted couplings considered there and here require a larger global $U(1)_{D}$ symmetry to forbid Majorana couplings of the dark fermion.} The behaviour of dark particles in the "freeze-in" scenario \cite{Hall:2009bx,Bernal:2017kxu} with different dark particle masses, dark sector couplings and RH neutrino masses was studied in \cite{Chianese:2018dsz, Becker:2018rve}. A general model was introduced but the masses and couplings of the two dark particles were assumed to be degenerate, in order to minimise the number of free parameters. Nevertheless, a successful dark matter relic density was shown to arise from the frozen-in particles, including an interesting region of parameter space involving superheavy dark matter particles which were dubbed ``FIMPzillas''.

In this paper, we revisit this proposal to achieve a minimal seesaw extension to simultaneously account for realistic neutrino mass and mixing, the baryon asymmetry of the Universe via leptogenesis and a viable dark matter relic density, in which two right-handed neutrinos are coupled to a dark Dirac fermion and complex scalar field, both charged under a global $U(1)_D$ symmetry. We again consider the Littlest Seesaw model which describes neutrino mass and mixing, but this time consider the requirement that the model also  accounts for leptogenesis, which has the effect of fixing the neutrino Yukawa couplings and right-handed neutrino masses. By again considering the freeze-in production mechanism of dark matter, we explore the parameter space of right-handed neutrino portal couplings and dark particle masses which give the correct dark matter relic abundance, focussing on the case of a superheavy Dirac fermion dark matter particle, with a mass around $10^{10}$ GeV, consistent with the leptogenesis constrained model. We show that such a FIMPzilla can provide a successful explanation of the dark matter relic abundance, with its production driven by neutrino Yukawa couplings over much of the parameter space, depending on the assumed dark particle masses, and the reheating temperature of the Universe.

We emphasise that, unlike the previous study in~\cite{Chianese:2018dsz}, the mass of RH neutrinos is here considered as fixed by the requirement of successful leptogenesis as in {\bf{Case A1}} in \cite{King:2018fqh}, which has the smallest $\Delta \chi^2$, and in this case the number of free parameters in the model is still four, the same as in \cite{Chianese:2018dsz}. This allows us to focus more on the relation between the remaining free parameters, namely the two RHN portal couplings and their dependence on the masses of the dark particles. The evolution of dark particle abundance will also be explored in detail. In addition to the free parameters in the model, the reheating temperature will be varied and treated as a free parameter in the cosmological model and its the effect on the dark couplings as well as its joint effect with the masses of dark particles will also be discussed.

The paper will be organised according to the following scheme. In section~\ref{sec:mod}, we briefly review of the general model linking the DM and the LS model. In sections~\ref{sec:Beqs} and~\ref{sec:BeqsH}, we derive the Boltzmann equations which are required to solve the dark matter production in the freeze-in scenario. In section~\ref{sec:res}, we show the numerical results, including the scanning the dark couplings, the abundance of dark particles and the interaction rates for different processes, the effect of changing reheating temperature. Finally, we summarise and conclude in section~\ref{sec:con}.

\section{The model\label{sec:mod}}

The model being considered here is an extension of the Littlest Seesaw model in which the two right-handed (RH) neutrinos $N_{Ri}$ are coupled to a dark Dirac fermion $\chi$ and a dark complex scalar $\phi$ in the dark sector (DS) through a Yukawa-like operator in the Lagrangian. The interactions between dark particles and the Standard Model particles are suppressed by a global $U(1)_{D}$ dark symmetry - the dark fields are charged and the other fields are neutral under the global $U(1)_{D}$ transformation. The detailed properties of the dark particles and the right-handed neutrinos under internal symmetries are shown in Tab.~\ref{tab:matter}. 
\begin{table}[t!]
\centering
\begin{tabular}{|c|c|c|c|}
\hline 
& $N_{Ri}$ & $\phi$ & $\chi$ \\ \hline \hline 
$SU(2)_L$ & {\bf 1} & {\bf 1} & {\bf 1} \\ \hline
$U(1)_Y$ & 0 & 0 & 0 \\ \hline  \hline
$U(1)_D$ & 0 & 1 & 1 \\ \hline
\end{tabular}
\caption{\label{tab:matter}New fields representations of the model, where $N_{Ri}$ are the two right-handed neutrinos, while $\phi$ and $\chi$ are a dark complex scalar and dark Dirac fermion, respectively, charged under a global 
$U(1)_{D}$ dark symmetry, but neutral under the electroweak $SU(2)_L\times U(1)_Y$ gauge symmetry.}
\end{table}
The total Lagrangian can be divided into four parts
\begin{equation}
\mathcal{L} = \mathcal{L}_{\rm SM} + \mathcal{L}_{\rm Seesaw} + \mathcal{L}_{\rm DS} + \mathcal{L}_{\rm N_Rportal}\,.
\label{eq:lag}
\end{equation}
Apart from the Standard Model (SM) sector, the remaining terms we consider are
\begin{eqnarray}
\mathcal{L}_{\rm seesaw} & = & - Y_{\alpha i} \overline{L_L}_\alpha \tilde{H} N_{Ri} - \frac12 M_{Rij}\overline{N^c_{Ri}} N_{Rj} + {\rm h.c.}\,, \label{eq:lagNS} \\
\mathcal{L}_{\rm DS} & = & \overline{\chi}\left(i \slashed{\partial} - m_\chi \right)\chi + \left|\partial_\mu \phi\right|^2 - m^2_\phi \left|\phi\right|^2 + V\left(\phi\right)\,, \label{eq:lagDS} \\
\mathcal{L}_{\rm N_Rportal} & = & y_{i} \phi \, \overline{\chi}N_{Ri} + {\rm h.c.} \,, \label{eq:lagPortal}
\end{eqnarray}
where ${L_{L\alpha}}$ are the three left-handed (LH) lepton doublets coupled to the two right-handed (RH) neutrinos $N_{Ri}$
($i=1,2$ and $\alpha=1,2,3$) and $\tilde{H} = i \tau_2 H^*$ with $H$ being the SM Higgs doublet
\begin{equation}
H = \left(\begin{array}{c}G^+ \\ \frac{v_{\rm SM} + h^0 + i G^0}{\sqrt2}\end{array}\right)\,.
\end{equation}
In Eq.~\eqref{eq:lagDS}, the quantities $m_\chi$ and $m_\phi$ are the masses of the dark Dirac fermion and the scalar, respectively. The term $V\left(\phi\right)$ is a general potential for the scalar field allowed by the $U(1)_{D}$ symmetry. We require that the scalar field does not acquire a v.e.v. so that the $U(1)_{D}$ symmetry is preserved. Finally, the right-handed neutrino portal defined in Eq.~\eqref{eq:lagPortal} allows the coupling of the visible and dark sectors, since $N_{Ri}$ is in the visible seesaw sector. Another allowed coupling between the two sectors is the so-called Higgs portal $y_{H\phi} \left|H\right|^2\left|\phi\right|^2$. However, in order to study the relation between neutrinos and dark matter, we focus on the right-handed neutrino portal while we take the coupling $y_{H\phi}$ to be negligible.\footnote{We have found that the Higgs portal $y_{H\phi}$ coupling has to be less $\mathcal{O}(10^{-10})$ in order to not play any role in the dark matter production. In particular, there is a threshold for the coupling $y_{H\phi}$ around $3.755\times10^{-10}$ in case of the highest possible allowed value for reheating temperature of the Universe. Below such a threshold, the right-handed neutrino portal couplings will keep the same magnitude as without the Higgs portal coupling. If $y_{H\phi}$ exceeds the threshold, the dark particles will be in general overproduced in freeze-in scenario and the current dark matter abundance cannot be achieved.}

The seesaw Lagrangian defines the coupling between neutrinos and the Higgs field, as well as the Majorana mass of the right-handed neutrinos. The spontaneous electroweak symmetry breaking gives the neutrinos a Dirac mass matrix 
\begin{equation}
m_D = \frac{v_{\rm SM}}{\sqrt 2} Y\,,
\label{mD1}
\end{equation}
which generates the effective left-handed Majorana neutrino mass matrix through type-I seesaw mechanism~\cite{Minkowski:1977sc,Yanagida:1979as,GellMann:1980vs,Schechter:1980gr,Mohapatra:1979ia,Mohapatra:1980yp}
\begin{equation}
m_\nu = -m_D M^{-1}_R {m_D}^T \,.
\label{eq:ssformula}
\end{equation}
In the Littlest Seesaw model~\cite{King:2013iva, Bjorkeroth:2014vha, King:2015dvf,Bjorkeroth:2015ora,Bjorkeroth:2015tsa,King:2016yvg,Ballett:2016yod,King:2018fqh}, the structure of the Yukawa matrix is set by the so-called constrained sequential dominance with $n=3$ (CSD(3)). The model has five free parameters only: two real couplings, $a$ and $b$, their relative phase $\eta$ and the two right-handed neutrino masses, $M_1$ and $M_2$. As shown in~\cite{King:2018fqh}, this very minimal framework is able at the same time to explain the measured neutrino mixing and mass parameters and to account for the baryon symmetry of the Universe through leptogenesis. In case of mass ordering $M_1 \ll M_2$, the neutrino Dirac mass matrix~\eqref{mD1} takes the form
\begin{equation}
m_D = \left(\begin{array}{lr} 0 &  b \, e^{i \frac{\eta}{2}} \\ a & 3b \, e^{i \frac{\eta}{2}} \\ a & b \, e^{i \frac{\eta}{2}} \end{array}\right) \,.
\label{mD2}
\end{equation}
in the basis where the right-handed neutrino Majorana mass matrix is diagonal
\begin{equation}
M_R =  \left(\begin{array}{cc} M_1 &  0 \\ 0 & M_2 \end{array}\right)
\end{equation}
These expressions for the Dirac mass matrix and right-handed neutrino Majorana mass matrix 
follow from a model of the kind discussed in~\cite{King:2013iva, Bjorkeroth:2014vha, King:2015dvf,Bjorkeroth:2015ora,Bjorkeroth:2015tsa,King:2016yvg,Ballett:2016yod,King:2018fqh}.
From the point of view of this paper, we shall follow the approach in~\cite{King:2018fqh} where the structure of these matrices
was simply assumed and the minimal number of parameters (four real parameters) were fitted using neutrino data and leptogenesis.
Thus the form of the Dirac mass matrix and right-handed neutrino Majorana mass matrix may be regarded as an ansatz capable of describing
the neutrino data and leptogenesis with a minimal number of parameters.
 As a result, the effective neutrino Majorana mass matrix for the left-handed neutrinos $\nu_L$ given by the seesaw mechanism is 
\begin{equation}
m_\nu = m_a \left(\begin{array}{ccc} 0 & 0 & 0 \\ 0 & 1 & 1 \\ 0 & 1 & 1 \end{array}\right) + m_b \, e^{i \eta} \left(\begin{array}{ccc} 1 & 3 & 1 \\ 3 & 9 & 3 \\ 1 & 3 & 1 \end{array}\right)\,.
\label{eq:numass}
\end{equation}
with
\begin{equation}
m_a = \frac{a^2}{M_1} \qquad {\rm and} \qquad m_b = \frac{b^2}{M_2} \,.
\label{mD3}
\end{equation}
Therefore, the Yukawa matrix in the first term of Eq.~\eqref{eq:lagNS} can be written as
\begin{equation}
Y = \sqrt{\frac{2 \,m_a\, M_1}{v^2_{SM}}} \left(\begin{array}{cc} 0 & 0 \\ 1 & 0 \\ 1 & 0\end{array}\right) + \sqrt{\frac{2 \,m_b\, M_2}{v^2_{SM}}} e^{i \frac{\eta}{2}} \left(\begin{array}{cc} 0 & 1 \\ 0 & 3  \\ 0 & 1\end{array}\right)\,.
\label{eq:yuk}
\end{equation}
In case of the other allowed mass ordering $M_2 \ll M_1$, the two columns of the Dirac mass matrix trade places. In this paper, we consider the first scenario with the mass ordering $M_1 \ll M_2$ since the other scenario can be recovered by simply switching the role of the two dark sector couplings. Hence, according to Ref.~\cite{King:2018fqh}, we take the following benchmark values for the right-handed neutrino Majorana masses
\begin{equation}
M_1=5.10\times 10^{10} \, \text{GeV}  \qquad \text{and}  \qquad M_2=3.28\times 10^{14} \, \text{GeV} \,,
\label{eq:RHmass}
\end{equation}
and the Dirac mass matrix in Eq.~\eqref{mD2} defined by
\begin{equation}
a = 1.42~{\rm GeV} \qquad \text{and}  \qquad b = 37.4~{\rm GeV} \qquad \text{and}  \qquad \eta = 2\pi/3 \,.
\end{equation}
These values have been obtained by fitting the active neutrino observables, requiring successful leptogenesis, and taking into account the effect of renormalization group equations under the assumption of a grand unification scale at $M_{\rm GUT} = 1.0 \times 10^{16}$~GeV.

The remaining free parameters of the Lagrangian in Eq.~\eqref{eq:lag} are therefore the two RH neutrino portal couplings $y_1$ and $y_2$ and the two masses $m_\chi$ and $m_\phi$. As will be discussed in the next sections, the requirement that one of the two particles in the dark sector plays the role of dark matter provides a constraint on the two DS couplings for a given choice of the masses.

\section{Dark Matter Production\label{sec:Beqs}}

The dependence of the particle yields on the temperature of the thermal bath is described by the Boltzmann equations. A general and useful form is~\cite{Kolb:1990vq}
\begin{eqnarray}
\mathcal{H}\,T\left(1+\frac{T}{3 g^\mathfrak{s}_*\left(T\right)}\frac{d g^\mathfrak{s}_*}{d T}\right)^{-1}\frac{d Y_i}{d T} &=& \sum_{kl} \left<\Gamma_{i\rightarrow kl}\right>Y_i^{\rm eq}\left(\frac{Y_i}{Y_i^{\rm eq}}-\frac{Y_k \, Y_l}{Y_k^{\rm eq}Y_l^{\rm eq}}\right)	\nonumber \\
&& - \sum_{jk} \left<\Gamma_{j\rightarrow ik}\right>Y_j^{\rm eq}\left(\frac{Y_j}{Y_j^{\rm eq}}-\frac{Y_i \, Y_k}{Y_i^{\rm eq}Y_k^{\rm eq}}\right)
\label{eq:Boltz} \\
&&+ \mathfrak{s} \sum_{jkl} \left<\sigma_{ij\rightarrow kl}\, v_{ij}\right> Y_i^{\rm eq}Y_j^{\rm eq}\left(\frac{Y_i \, Y_j}{Y_i^{\rm eq}Y_j^{\rm eq}}-\frac{Y_k \, Y_l}{Y_k^{\rm eq}Y_l^{\rm eq}}\right) \,. \nonumber
\end{eqnarray}
In the left-hand side of Eq.\eqref{eq:Boltz}, the quantities $\mathcal{H}$ and $\mathfrak{s}$ are the Hubble parameter and the entropy density of the thermal bath\footnote{The symbol $ \mathfrak{s}$ is used to represent the entropy density while the symbol $s$ denotes the Mandelstam variable.} defined by
\begin{equation}
\mathcal{H} = 1.66 \sqrt{g_*\left(T\right)}\frac{T^2}{M_{\rm Planck}} \qquad{\rm and}\qquad \mathfrak{s} = \frac{2\pi^2}{45} g^\mathfrak{s}_*\left(T\right) T^3 \,,
\label{eq:hubble}
\end{equation}
where the Planck mass $M_{\rm Planck}$ is $1.22\times10^{19}$ GeV, and $g_*$ and $g^\mathfrak{s}_*$ are the degrees of freedom of the relativistic species in the thermal bath. At high temperature, we have $g_* = g^\mathfrak{s}_* = 106.75$ according to the SM particle content. In Eq.~\eqref{eq:Boltz}, the yield of particles $i$ at the thermal equilibrium is denoted by $Y^{\rm eq}_i$, which takes the expression 
\begin{equation}
Y^{\rm eq}_{i} \equiv \frac{n^{\rm eq}_{i}}{\mathfrak{s}}\qquad{\rm with}\qquad n^{\rm eq}_{i} = \frac{g_i \, m_i^2 \, T}{2\pi^2} K_2\left(\frac{m_i}{T}\right)\,,
\label{eq:yeq}
\end{equation}
with $m_i$ the mass and $g_i$ the internal degrees of freedom of the particle $i$, and $K_2$ the order-2 modified Bessel function of the second kind. In Eq.~\eqref{eq:Boltz}, the summations cover all the possible decay processes with particle $i$ as the decaying particle (the first term) and as the decay product (the second term) and all the possible scattering processes where particle $i$ appears in the initial state (the third term). In particular, the thermally averaged decay width is given by
\begin{equation}
\left< \Gamma_{i\rightarrow kl} \right> = \frac{K_1\left(m_i/T\right)}{K_2\left(m_i/T\right)}\Gamma_{i\rightarrow kl}\,,
\end{equation}
where $K_1$ is the order-1 modified Bessel function of the second kind. The thermally averaged cross section is~\cite{Edsjo:1997bg}
\begin{equation}
\left<\sigma_{ij\rightarrow kl}\, v_{ij}\right> = \frac{1}{n^{\rm eq}_{i}\, n^{\rm eq}_{j}}\frac{g_i \, g_j}{S_{kl}}\frac{T}{512\pi^6} \int_{\left(m_i+m_j\right)^2}^\infty ds \, \frac{p_{ij} \, p_{kl} \,K_1\left(\sqrt{s}/T\right)}{\sqrt{s}} \int  \overline{\left|\mathcal{M}\right|^2}_{ij\rightarrow kl} \, d\Omega\,.
\label{avgsection}
\end{equation}
with $s$ the square of the centre-of-mass energy, $S_{kl}$ the symmetry factor, and $p_{ij}$ ($p_{kl}$) the initial (final) centre-of-mass momentum.

In the present paper, we investigate very heavy dark matter particles, meaning that the masses of particles in the dark sector are much larger than the electroweak scale. In this case, the standard freeze-out production mechanism~\cite{Gondolo:1990dk} is not viable since it would lead to an over-production of dark matter particles due to the unitarity limit on the cross-section~\cite{Griest:1989wd}. Unitarity indeed sets an upper bound of about 100~TeV to the mass of dark matter particles in order to be thermally produced. On the other hand, very heavy dark matter particles can be non-thermally produced through the so-called freeze-in production~\cite{Hall:2009bx}. In this case, the dark sector couplings are assumed to be very small and, consequently, the corresponding interaction rates are not able to bring the dark sector into thermal equilibrium with the thermal bath. In the freeze-in framework, the dark particles are therefore produced gradually through very weak interactions from the thermal bath. Different from the freeze-out mechanism where the thermal equilibrium erases any information on the initial settings, the freeze-in mechanism is strongly dependent on the initial conditions assumed to solve the Boltzmann equations. Typically, one considers a negligible abundance of particles in the dark sector at the end of inflation. These initial conditions are imposed at the reheating temperature of the Universe, hereafter denoted as $T_{\rm RH}$. When the reheating temperature is higher than DM mass, the dark couplings have to be small enough to suppress the dark particle production. However, as the reheating temperature becomes lower, the dark couplings need to be larger to compensate for the shorter cosmological period available for DM production. After freezing-in, the total DM relic abundance is given by
\begin{equation}
\Omega_{\rm DM}h^2 = \frac{\rho_{\rm DM,0}}{\rho_{\rm crit}/h^2}\,,
\label{eq:omegaPRE}
\end{equation}
where $\rho_{\rm DM,0}$ is the dark matter energy density of today and $\rho_{\rm crit}/h^2 = 1.054 \times10^{-5} {\rm GeV \, cm^{-3}}$ is the critical density \cite{Patrignani:2016xqp}. 
Since the aim of this paper is to make a model consistent with the current data,
the left-hand side of Eq.\eqref{eq:omegaPRE} is set to be the measured value provided by the Planck Collaboration at 68\% C.L. \cite{Ade:2015xua}:
\begin{equation}
\left.\Omega_{\rm DM}h^2\right|_{\rm obs} = 0.1188 \pm 0.0010\,.
\label{eq:omegaOBS}
\end{equation}

In principle, five coupled Boltzmann equations should be solved to obtain a full description of the DM relic abundance: two for the two right-handed neutrinos, two for the DM particles and one describing the evolution of baryon asymmetry of the Universe. However, considering the masses of the right-handed neutrinos are much larger than the electroweak scale (see Eq.~\eqref{eq:RHmass}), the neutrino Yukawa couplings given in Eq.~\eqref{eq:yuk} are large enough to bring the neutrino sector into thermal equilibrium with the thermal bath. Hence, the yields of the right-handed neutrinos follow the equilibrium distribution defined in Eq.~\eqref{eq:yeq}. Moreover, we note that other terms induced by the dark sector interactions in the Boltzmann equations of the right-handed neutrinos and in the one for the baryon asymmetry are highly suppressed according to the freeze-in production mechanism. The suppression is either due the smallness of the dark sector couplings $y_i$ or by the smallness of the yields of dark sector particles, i.e. $Y_\phi << Y^{\rm eq}_\phi$ and $Y_\chi << Y^{\rm eq}_\chi$.\footnote{We also note that, for the large dark matter mass considered in this paper, the yield of dark sector particles is in general many orders of magnitude smaller than the one of the baryon asymmetry of the universe, i.e. $Y_B = \left(0.87 \pm 0.01 \right) \times 10^{-10}$~\cite{Ade:2015xua}. This further justifies the validity of the standard leptognesis paradigm as studied in Ref.~\cite{King:2018fqh}.} Therefore, only the two coupled Boltzmann equations for the dark particles are remaining to be solved.

As discussed in Ref.~\cite{Chianese:2018dsz}, the processes entering in the Boltzmann equations are defined by the mass ordering between the dark sector particles and the right-handed neutrinos. In Ref. \cite{Chianese:2018dsz}, the cases in which $m_\phi \ge M_R \,, m_\chi$ and $M_R \ge m_\phi \,, m_\chi$ are discussed based on the assumption that the right-handed neutrinos have the same mass $M_R$. However, since we are now considering non-degenerate right-handed neutrino masses, there is an additional scenario where the mass of the scalar dark particle lies between $M_1$ and $M_2$. Thus there are three ordering types in total: 
\begin{itemize}
\item {\bf Heavy scalar mass (ordering type A):} $m_\phi > M_1,\, M_2,\, m_\chi  $;
\item {\bf Intermediate scalar mass (ordering type B):} $M_2  > m_\phi > M_1,\, m_\chi $;
\item {\bf Light scalar mass (ordering type C):} $M_2  > M_1 > m_\phi > m_\chi$
\end{itemize}
In all the cases, the dark Dirac fermion $\chi$ is the lightest dark particle and therefore is stable.
This classification of mass ordering is based on the mass of $\phi$ particles, which sets the decay processes of the scalar dark particle and the RH neutrinos. There are exceptional cases in each ordering type where the mass of the particles are close to each other and the two-body decay is suppressed, including 
\begin{itemize}
\item$m_\phi<m_\chi + M_2$ in ordering type A,
\item$M_2  < m_\phi + m_\chi$ or $m_\phi < m_\chi + M_1$ in ordering type B,
\item$M_1 < m_\phi + m_\chi$ in ordering type C.
\end{itemize}
In this paper, we will focus on ordering type A for which we expect the scattering processes involving the Yukawa couplings drive the DM production as discussed in Ref.~\cite{Chianese:2018dsz}. In this case, $\phi$ can decay into $\chi$ and a RH neutrino through two-body decay. We conclude this section by commenting on the other orderings. In ordering type B, the DM production is expected to be dominated by the two-body decay of the heaviest RH neutrino into $\chi$ and $\phi$, which then decays into $\chi$ through the two-body decay involving the lightest RH neutrino. The former decay is set by the dark coupling $y_2$ while the latter is weighted by $y_1$. Other scattering processes are instead suppressed by large powers of these couplings and, consequently, the Yukawa neutrino interactions are expected to provide a subdominant contribution in most of the DM parameter space. This means that the DM production is mainly driven by the dark sector. Finally, in ordering type C, the two-body decay of $\phi$ is totally suppressed while both of the RH neutrinos can decay into $\chi$ and $\phi$ through two-body decay. As studied in Ref.~\cite{Chianese:2018dsz}, this scenario is strongly constrained by cosmological observations and indirect DM searches due to the three-body decay of $\phi$ particles occurring after Big Bang Nucleosynthesis.

\section{Boltzmann equations for heavy scalar mass\label{sec:BeqsH}}

\begin{figure}[t!]
\begin{center}
\subfigure[Dark Sector scatterings]{\begin{fmffile}{DS1}
\fmfframe(18,18)(18,18){
\begin{fmfgraph*}(80,45)
\fmflabel{$\phi^*$}{i1}
\fmflabel{$\phi$}{i2}
\fmflabel{$n_j$}{o1}
\fmflabel{$n_i$}{o2}
\fmfv{label=$y_{i}$}{v1}
\fmfv{label=$y_{j}$}{v2}
\fmfleft{i1,i2}
\fmfright{o1,o2}
\fmf{dashes}{i2,v1}
\fmf{plain}{v1,o2}
\fmf{dashes}{i1,v2}
\fmf{plain}{v2,o1}
\fmf{plain,label=$\chi$,tension=0}{v1,v2}
\fmfdotn{v}{2}
\end{fmfgraph*}}
\end{fmffile}
\begin{fmffile}{DS2}
\fmfframe(18,18)(18,18){
\begin{fmfgraph*}(80,45)
\fmflabel{$\overline{\chi}$}{i1}
\fmflabel{$\chi$}{i2}
\fmflabel{$n_j$}{o1}
\fmflabel{$n_i$}{o2}
\fmfv{label=$y_{i}$}{v1}
\fmfv{label=$y_{j}$}{v2}
\fmfleft{i1,i2}
\fmfright{o1,o2}
\fmf{plain}{i2,v1}
\fmf{plain}{v1,o2}
\fmf{plain}{i1,v2}
\fmf{plain}{v2,o1}
\fmf{dashes,label=$\phi$,tension=0}{v1,v2}
\fmfdotn{v}{2}
\end{fmfgraph*}}
\end{fmffile}}
\subfigure[Neutrino Yukawa scatterings]{\begin{fmffile}{NS1}
\fmfframe(23,18)(18,18){
\begin{fmfgraph*}(80,45)
\fmflabel{$\phi,\phi^*$}{i1}
\fmflabel{$\overline{\chi},\chi$}{i2}
\fmflabel{$\nu_i,\nu_i,\ell_i^\pm$}{o2}
\fmflabel{$h^0,G^0,G^\mp$}{o1}
\fmfv{label=$y_{j}$}{v1}
\fmfv{label=$y_\nu$}{v2}
\fmfleft{i1,i2}
\fmfright{o1,o2}
\fmf{plain}{i2,v1}
\fmf{dashes}{i1,v1}
\fmf{dashes}{v2,o1}
\fmf{plain}{v2,o2}
\fmf{plain,label=$n_j$}{v1,v2}
\fmfdotn{v}{2}
\end{fmfgraph*}}
\end{fmffile}}\qquad\qquad
\end{center}
\caption{\label{fig:Feyn1}Dominant scattering processes responsible for DM production in case of heavy dark scalar (ordering type A). Here $n_i$ represent the heavy neutrino mass eigenstates (dominantly from the right-handed neutrinos $N_R$) while $\nu_i$ represent the light neutrino mass eigenstates (dominantly from $\nu_L$ in the doublet $L_L$). }
\end{figure}
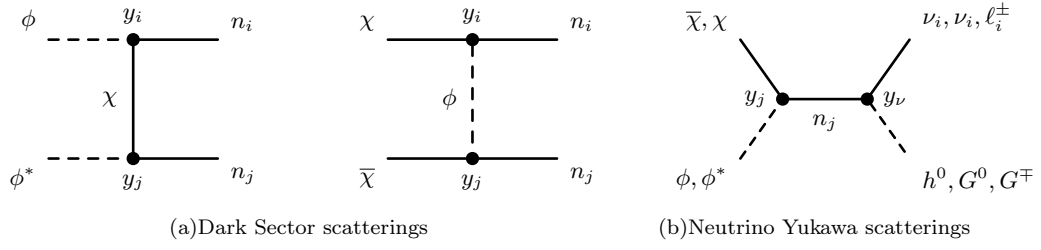

In ordering type A, three kinds of scattering process are involved in the Boltzmann equations for the two dark species: $\phi\phi^*$ scattering, $\chi\overline{\chi}$ scattering and $\phi\overline{\chi}$ ($\phi^*\chi$) scattering
(see Fig.~\ref{fig:Feyn1}). In addition, the two-body decay of $\phi$ also contributes to the Boltzmann equations.
These equations are given by
\begin{eqnarray}
\mathcal{H}\,T\left(1+\frac{T}{3 g^\mathfrak{s}_*\left(T\right)}\frac{d g^\mathfrak{s}_*}{d T}\right)^{-1} \frac{d Y_\phi}{d T} & = & 
- \mathfrak{s} \left<\sigma\, v\right>_{\phi\phi}^{\rm DS} \left({Y_\phi^{\rm eq}}\right)^2 
- \mathfrak{s} \left<\sigma\, v\right>^{\rm \nu-Yukawa}_{\chi\phi} Y_\phi^{\rm eq}Y_\chi^{\rm eq} \nonumber \\
& & + \left<\Gamma_{\phi}\right>\left(Y_\phi-\frac{Y_\phi^{\rm eq}}{Y_\chi^{\rm eq}}Y_\chi\right)  \,, \label{eq:phia} \\
\mathcal{H}\,T\left(1+\frac{T}{3 g^\mathfrak{s}_*\left(T\right)}\frac{d g^\mathfrak{s}_*}{d T}\right)^{-1} \frac{d Y_\chi}{d T} & = & 
- \mathfrak{s}  \left<\sigma\, v\right>_{\chi\chi}^{\rm DS} \left({Y_\chi^{\rm eq}}\right)^2 
- \mathfrak{s} \left<\sigma\, v\right>^{\rm \nu-Yukawa}_{\chi\phi} Y_\phi^{\rm eq}Y_\chi^{\rm eq}  \nonumber \\
& & -  \left<\Gamma_{\phi}\right>\left(Y_\phi-\frac{Y_\phi^{\rm eq}}{Y_\chi^{\rm eq}}Y_\chi\right) \,, \label{eq:chia}
\end{eqnarray}
where the right-handed neutrinos $N_{Ri}$ are taken to be in thermal equilibrium. In the above expressions, we assume that $Y_{\chi,\phi} \ll Y_{\chi,\phi}^{\rm eq}$ according to the freeze-in production paradigm and the yields of all the other particles are at thermal equilibrium. Moreover, we do not consider the other scattering processes with right-handed neutrinos that are suppressed by the active-sterile neutrino mixing $\theta\equiv m_D M^{-1}_R$~\cite{Buchmuller:1990vh,Pilaftsis:1991ug}. All the terms in Eq.s~\eqref{eq:phia} and~\eqref{eq:chia} are given by:\footnote{The detailed expressions for the scattering amplitudes are reported in the Appendix.} 

\begin{itemize}
\item {\bf Dark Sector scatterings:} 
In the Boltzmann equations, the first terms are $\phi\phi^*$ scatterings and $\chi\overline{\chi}$ scatterings for which
\begin{eqnarray}
\left<\sigma\, v\right>_{\phi\phi}^{\rm DS} & = & \sum_{i,j=1,2} \left<\sigma_{\phi\phi^*\rightarrow n_i n_j}\, v\right> \,, \\
\left<\sigma\, v\right>_{\chi\chi}^{\rm DS} & = & \sum_{i,j=1,2} \left<\sigma_{\chi\overline{\chi}\rightarrow n_i n_j}\, v\right> \,.
\end{eqnarray}
The amplitude covers process with all possible combinations of outgoing RH neutrinos. 
The amplitudes are proportional to $y_{1}^4$ for $N_{R1} N_{R1}$ final state, $y_{2}^4$ for $N_{R2} N_{R2}$ final state 
and $y_{1}^2 y_{2}^2$ for $N_{R1} N_{R2}$ final state.

\item {\bf Neutrino Yukawa scatterings:} 
The second term in both of the equations is the scattering of $\phi^*$($\phi$) and $\chi$($\overline{\chi}$), 
with standard model particles in the final state.
The outgoing particles could be LH neutrino and Higgs bosons or neutral Goldstone bosons, or charged leptons and charged Goldstone bosons:
\begin{equation}
\left<\sigma\, v\right>^{\rm \nu-Yukawa}_{\chi\phi} = \sum_{i=1}^3 \left[\left<\sigma_{\chi \phi\rightarrow \nu_i h^0}\, v\right> + \left<\sigma_{\chi \phi\rightarrow \nu_i G^0}\, v\right> + \left<\sigma_{\chi \phi\rightarrow \ell^\pm_i G^\mp}\, v\right>\right]\,.
\end{equation}
The amplitude of such process would be proportional to $y_{i}^2 y_\nu^2$, 
where $i=1\,,2$ depending on the intermediate right-handed neutrino 
and $y_\nu = \left(U^\dagger_\nu Y\right)_{ij}/\sqrt2 \,\text{or}\, Y_{ij}$ depending on the type of the outgoing lepton. The quantity $U_\nu$ is the Pontecorvo–Maki–Nakagawa–Sakata matrix describing neutrino oscillations.

\item {\bf Scalar decay:} 
The last terms are related to the two-body decays of $\phi$ particles into $\chi$ and right-handed neutrino, which are kinematically allowed if $m_\phi > m_\chi + M_i$. The corresponding partial decay widths are proportional to $y_{i}^2$ and take the expression
\begin{equation}
\Gamma_{\phi \rightarrow \chi n_i}^{\rm 2-body} = \frac{y_{i}^2\,m_\phi}{16 \pi}\left(1-\frac{m_\chi^2}{m_\phi^2}-\frac{M_{i}^2}{m_\phi^2}\right)\sqrt{\lambda\left(1,\frac{m_\chi^2}{m_\phi^2},\frac{M_i^2}{m_\phi^2}\right)}
\end{equation}
where $\lambda$ is the Kallen function. The total decay width is the sum of the two widths:
\begin{equation}
\Gamma_{\phi}^{\rm 2-body} = \sum_{i=1}^2 \Gamma_{\phi \rightarrow \chi n_i}^{\rm 2-body} \,.
\label{eq:totdecay}
\end{equation}
When the masses of the two kinds of dark particles and the heaviest RH neutrino are close to each other, there could be special circumstances in which $M_2 + m_\chi > m_\phi$. This implies that the mass of the scalar dark particle is not large enough to achieve the threshold of the two-body decay. In that case, the two-body decay process of $\phi$ into $N_{R2}$ and $\chi$ is kinematically suppressed but $\phi$  can still decay into $N_{R1}$ and $\chi$. Therefore the scalar dark particle still remains unstable and the single DM scenario will be kept.
\end{itemize}

The total yield of the dark matter can be described by the sum of the two Boltzmann equations: 
\begin{eqnarray}
\mathcal{H}\,T\left(1+\frac{T}{3 g^\mathfrak{s}_*\left(T\right)}\frac{d g^\mathfrak{s}_*}{d T}\right)^{-1} \frac{d Y_{\rm DM}}{d T} & = & - \mathfrak{s} \left<\sigma\, v\right>_{\phi\phi}^{\rm DS} \left({Y_\phi^{\rm eq}}\right)^2 - \mathfrak{s} \left<\sigma\, v\right>_{\chi\chi}^{\rm DS}  \left({Y_\chi^{\rm eq}}\right)^2 \nonumber \\
& & - 2 \, \mathfrak{s} \left<\sigma\, v\right>^{\rm \nu-Yukawa}_{\chi\phi} Y_\phi^{\rm eq}Y_\chi^{\rm eq}\,. \label{eq:Beq}
\end{eqnarray}
This differential equation can be easily integrated providing the today's DM yield at $T=0$, denoted as $Y_{\rm DM,0}$. Such a quantity is given by the sum of different contributions related to the dark sector couplings:
\begin{equation}
Y_{\rm DM,0} = Y^{\rm DS}_{n_1 n_1} + Y^{\rm DS}_{n_2 n_2} + Y^{\rm DS}_{n_1 n_2} 
+ 2\,Y^{\rm \nu-Yukawa}_{n_1} + 2\,Y^{\rm \nu-Yukawa}_{n_2 } + 2\,Y^{\rm \nu-Yukawa}_{\rm int}\,,
\label{eq:DMyield}
\end{equation}
where
\begin{eqnarray}
Y^{\rm DS}_{n_i n_j} &  = & \int_{0}^{T_{\rm RH}} dT \, \frac{\mathfrak{s}}{\mathcal{H}\,T} 
\left(1+\frac{T}{3 g^\mathfrak{s}_*\left(T\right)}\frac{d g^\mathfrak{s}_*}{d T}\right) \times \nonumber\\&&
\left( \left<\sigma_{\phi\phi^*\rightarrow n_i n_j} \, v\right> \left({Y_\phi^{\rm eq}} \right)^2 
+ \left<\sigma_{\chi\overline{\chi}\rightarrow n_i n_,} \, v\right> \left({Y_\chi^{\rm eq}} \right)^2\right)\,,
\end{eqnarray}
and $Y^{\rm \nu-Yukawa}_{n_i}$ refers to the contribution from Yukawa $\chi\phi$ scattering mediated by $N_{Ri}$  and $Y^{\rm \nu-Yukawa}_{\rm int}$ refers to the interference contribution (see the Appendix). Here, we have assumed negligible the yields of dark particles as an initial condition at the initial temperature set by $T_{\rm RH}$. The reheating temperature is constrained from above by the limit on tensor modes $r < 0.056$ set by Planck~\cite{Akrami:2018odb} and BICEP-Keck~\cite{Ade:2018gkx} at 95\% CL. In particular, we have
\begin{equation}
T_{\rm RH} \leq T_{\rm RH}^{\rm max} \simeq  6.5 \times 10^{15}~{\rm GeV}\,.
\end{equation}
If not explicitly stated, the reheating temperature is assumed to coincide with its maximum allowed value. It is worth observing that, for large values of the reheating temperature, very heavy dark matter particles can in general be produced by gravity-mediated interactions~\cite{Chung:2004nh,Garny:2015sjg,Garny:2017kha,Bernal:2018qlk}. This production mechanism depends in general on the scenario of inflation and reheating. A detailed analysis of such production mechanism is beyond the scope of the present paper and left to future work. However, we note that gravity production might be suppressed in case of non-instantaneous reheating or low reheating temperatures~\cite{Garny:2015sjg}. The latter case is discussed in the last part of the present paper.

We can rewrite Eq.~\eqref{eq:DMyield} in a way that one can explicitly see how it depends on the right-handed neutrino portal couplings $y_{1}$ and $y_{2}$: 
\begin{equation}
Y_{\rm DM,0} =\, y_{1}^4\,\tilde{Y}^{\rm DS}_{n_1 n_1} + y_{2}^4\,\tilde{Y}^{\rm DS}_{n_2 n_2} 
+ y_{1}^2y_{2}^2\,\tilde{Y}^{\rm DS}_{n_1 n_2} 
+ y_{1}^2\,\tilde{Y}^{\rm \nu-Yukawa}_{n_1} +  y_{2}^2\,\tilde{Y}^{\rm \nu-Yukawa}_{n_2}
+ y_{1}y_{2}\,\tilde{Y}^{\rm \nu-Yukawa}_{int}\,,
\label{eq:DMyieldYDSA}
\end{equation}
where the quantities $\tilde{Y}$ are suitably defined by extracting $y_{1}$ and $y_{2}$ from the corresponding expressions. According to Eq.~\eqref{eq:omegaPRE}, the total DM relic abundance predicted by the model takes the form
\begin{equation}
\Omega_{\rm DM}h^2 =  \frac{2 \, \mathfrak{s}_0 \, m_\chi  \,Y_{\rm DM,0}}{\rho_{\rm crit}/h^2}  \,,
\label{eq:omegaPREA}
\end{equation}
where $\mathfrak{s}_0=2891.2\,{\rm cm^3}$ is today’s entropy density \cite{Patrignani:2016xqp}, and the factor of 2 takes into account the contribution of DM anti-particles. Requiring that this expression matches the experimental value reported in Eq.~\eqref{eq:omegaOBS} provides constraints on the dark sector couplings $y_1$ and $y_2$ for a given choice of $m_\chi$ and $m_\phi$.

\section{Results\label{sec:res}}

In this section we collect the main results of the present analysis. We firstly report the allowed values for the DS couplings required to match the DM relic abundance. Then, we discuss the complete behavior of the yields in the dark sector after solving the Boltzmann equations for some benchmark cases. Lastly, we examine how the DS couplings and dark matter production depends on the reheating temperature of Universe and report the allowed parameters space through a full numerical scan of the model.

\subsection{Scanning the dark sector couplings $y_{1}$ and $y_{2}$\label{sec:couplings}}
\begin{figure}[t!]
\begin{center}
\subfigure[]{\includegraphics[width=0.48\textwidth]{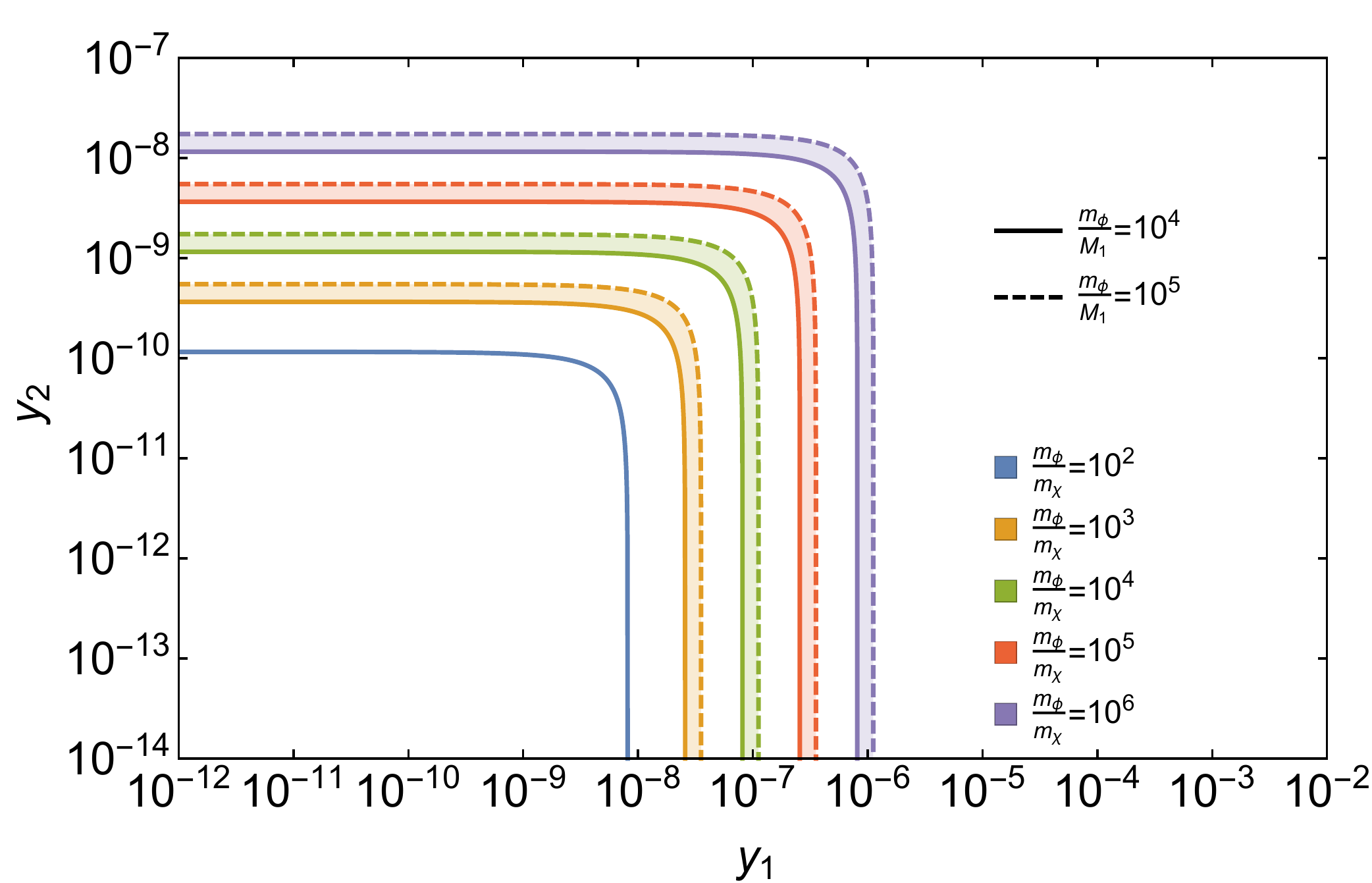}}
\subfigure[]{\includegraphics[width=0.48\textwidth]{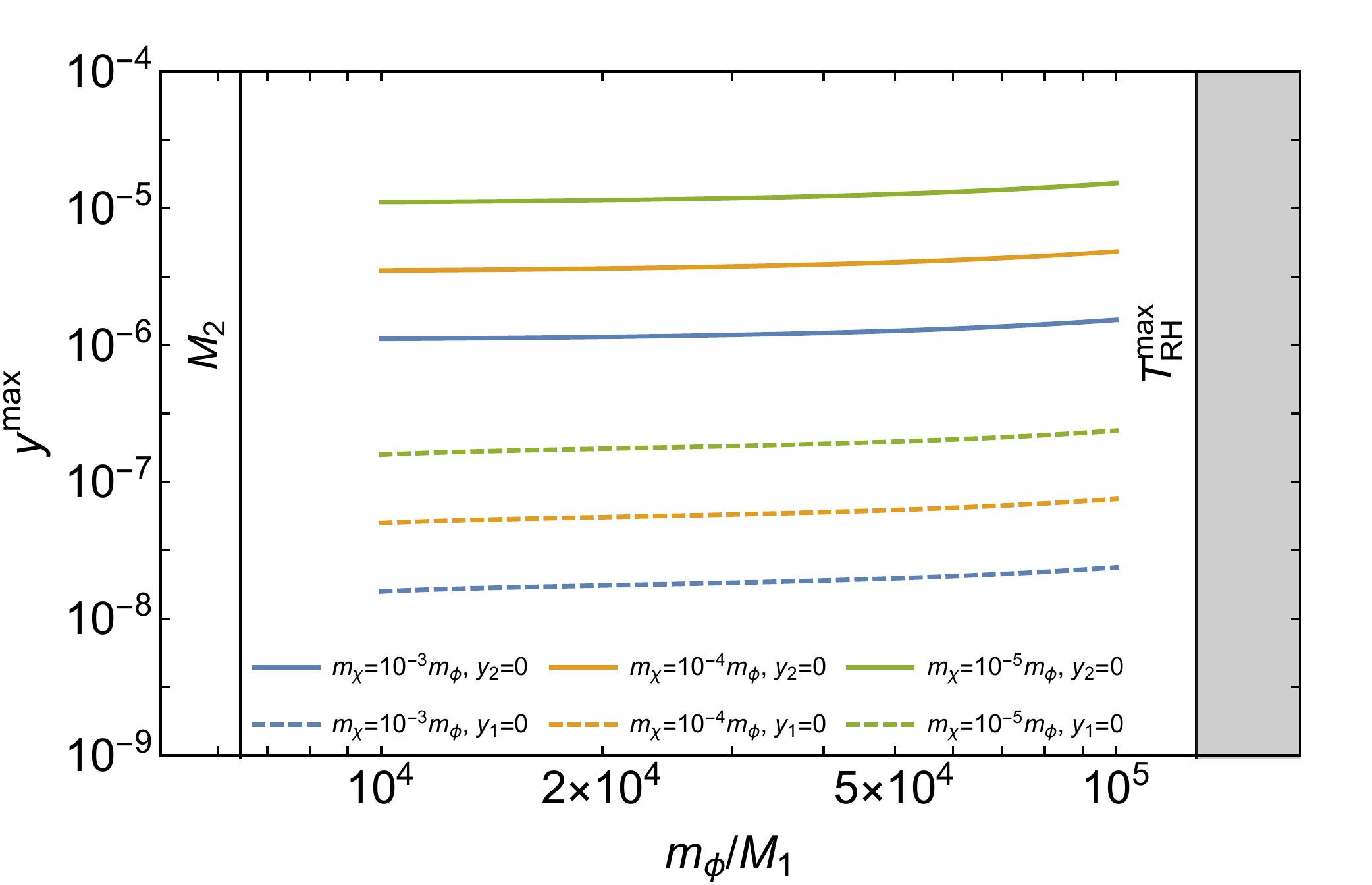}}
\caption{\label{fig:couplings}
(a) The dependence of the RHN portal couplings $y_{2}$ on $y_{1}$ for different ratios of the dark particle masses, required to achieve the correct dark matter relic abundance.
(b) The dependence of the maximum value of the dark sector couplings, $y_1$ (solid lines) and $y_2$ (dashed lines), as a function of the mass of $\phi$ in units of $M_1$. Different colors represent different ratios $m_\phi / m_\chi$. We recall that $M_1 = 5.10 \times 10^{10}~\mathrm{GeV}$, $M_2 = 3.28 \times 10^{14}~\mathrm{GeV}$ and $T_{\rm RH}^{\rm max} \simeq  6.5 \times 10^{15}~\mathrm{GeV}$.}
\end{center}
\end{figure}

By fixing the mass of the two dark particles, we can get a relation between the RH neutrino portal couplings, $y_{1}$ on $y_{2}$, accounting for the correct DM relic abundance. This relation is shown in Fig.~\ref{fig:couplings}(a). It can be observed that both $y_{1}$ and $y_{2}$ have an upper limit, which is achieved when the other one is zero. In other words, the maximum value for one of the two couplings is obtained when the other RH neutrino is completely decoupled from the dark sector. Such maximum values are determined by the following simplified forms of Eq.~\eqref{eq:DMyieldYDSA}:
\begin{equation}
\begin{aligned}
Y_{\rm DM,0} =\, (y^{\rm max}_{1})^4\,\tilde{Y}^{\rm DS}_{n_1 n_1} + (y^{\rm max}_{1})^2\,\tilde{Y}^{\rm \nu-Yukawa}_{n_1} 
\quad\text{and}\quad
Y_{\rm DM,0} =\, (y^{\rm max}_{2})^4\,\tilde{Y}^{\rm DS}_{n_2 n_2} + (y^{\rm max}_{2})^2\,\tilde{Y}^{\rm \nu-Yukawa}_{n_2}\,.
\label{eq:DMyieldYDSA2}
\end{aligned}
\end{equation}
When the mass of $\chi$ is fixed, the $y_2 - y_1$ curve moves upward and rightward as the mass of $\phi$ increases, which means the dark sector interactions have to be stronger in order to fulfil today's dark matter yield given by Eq.~\eqref{eq:omegaPREA}.  Another interesting fact is that the values of the dark sector couplings are nearly independent of the individual masses of dark particles but rather they depend on their ratio. In Fig.~\ref{fig:couplings}(a), different colors refer to different ratio of DS masses while the solid and dashed lines represent different values of the scalar mass $m_\phi$. Therefore, the shaded colored regions cover the slight dependence on the individual mass $m_\phi$. Such feeble dependence is highlighted in Fig.~\ref{fig:couplings}(b): the maximum values of the  DS couplings almost remain constant when $m_\phi /m_\chi$ is fixed. This is in agreement with previous results discussed in Ref.~\cite{Chianese:2018dsz}. It is useful to give some analytical insight into such numerical results. As will be discussed later, for high reheating temperatures, the neutrino Yukawa interactions dominate the DM production. In this case, the Boltzmann equation~\eqref{eq:Beq} can be further simplified as 
\begin{equation}
\frac{d Y_{\rm DM}}{d T} = - \frac{2 \, \mathfrak{s}}{\mathcal{H}\,T} \left(1+\frac{T}{3 g^\mathfrak{s}_*\left(T\right)}\frac{d g^\mathfrak{s}_*}{d T}\right) \left[ \left<\sigma\, v\right>^{\rm \nu-Yukawa}_{\chi\phi} Y_\phi^{\rm eq}Y_\chi^{\rm eq}\right]\,. \label{eq:Beq}
\end{equation}
In the limit of $m_\phi \gg m_\chi$, the term in the squared parenthesis does only depends on $m_\phi$ and $M_i$, as been seen by plugging Eq.~\eqref{avgsection} and the expressions for the scattering amplitudes reported in the Appendix. This implies that the final yield of DM particles should behave as $Y_{\rm DM,0} \propto y_i^2 \left(m_\phi / M_i \right)^\alpha$, being the yield a dimensionless quantity and $m_\phi$ and $M_i$ the only mass scales in these scattering processes. The differential equation~\eqref{eq:Beq} cannot be solved analytically. However, the numerical scan of the DM parameter space points out that the scaling power is $\alpha=-1$, as also partially shown in Fig.~\ref{fig:couplings}. Therefore, for a given value of the right-handed neutrino mass, by taking the observed value of the DM relic abundance~\eqref{eq:omegaOBS} and inverting Eq.~\eqref{eq:omegaPREA}, we get  $(y^{\rm max}_i)^2 \sim m_\phi / m_\chi$. Such a result is also justified by the fact that $\Omega_{\rm DM}$ is a dimensionless quantity.

\subsection{Dark sector yields and the interaction rates\label{sec:DandR}}
\begin{figure}[t!]
\begin{center}
\subfigure[]{\includegraphics[width=0.48\textwidth]{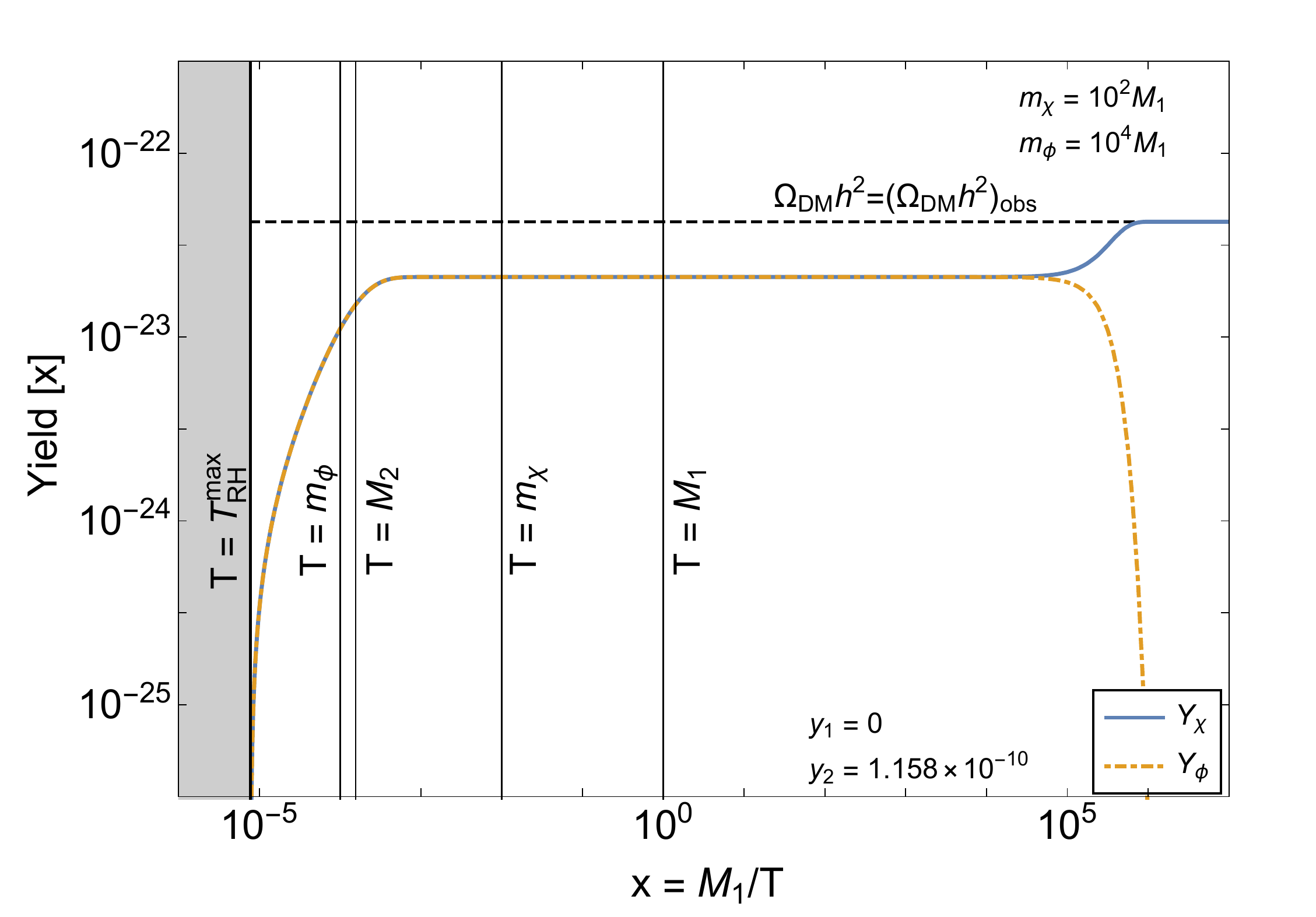}}
\hskip3.mm
\subfigure[]{\includegraphics[width=0.48\textwidth]{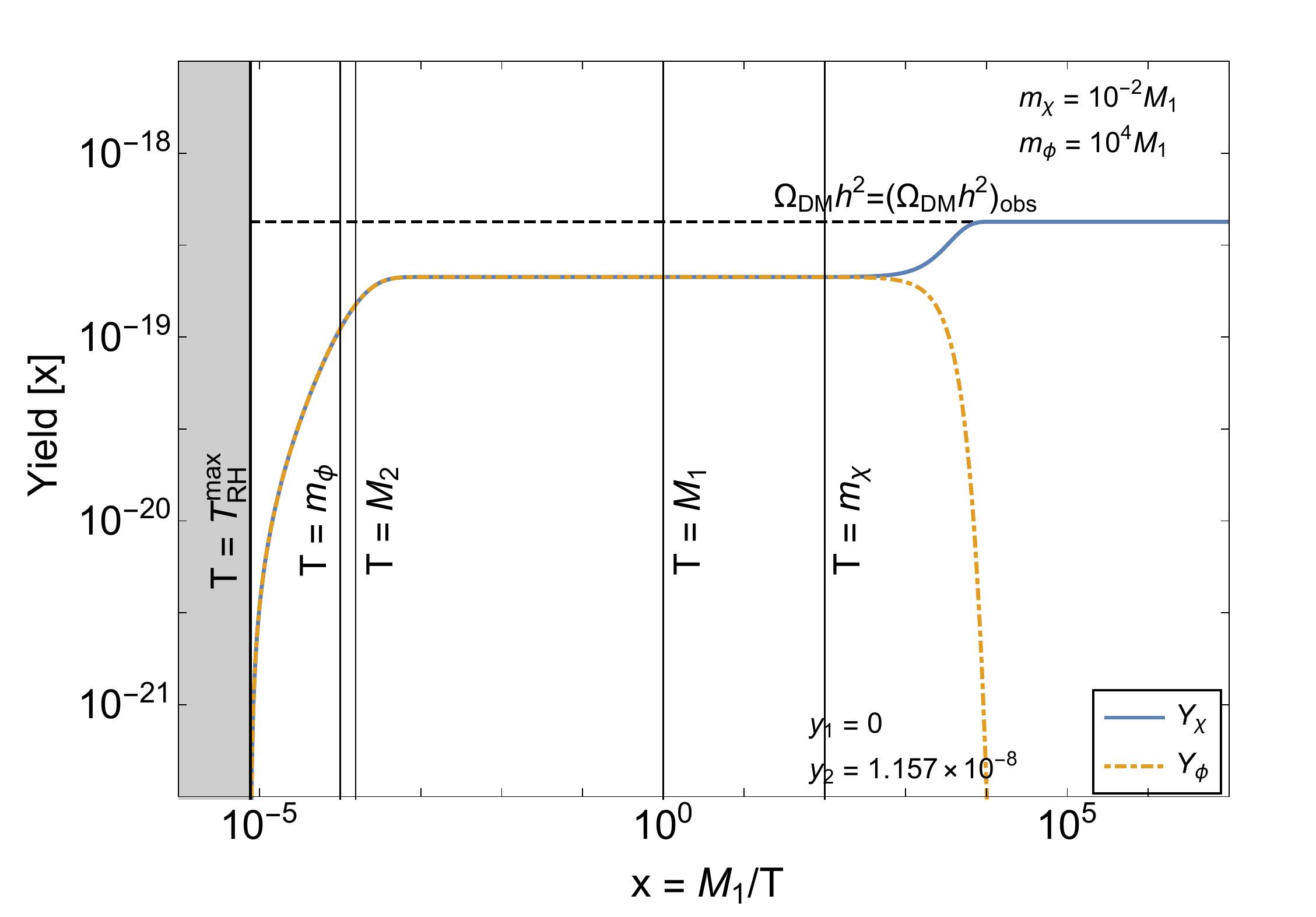}}
\caption{\label{fig:Yield}The yields of dark particles as a function of the auxiliary variable $x = M_1 / T$ for different masses of dark particles. The reheating temperature coincides with its maximum allowed value, at which we set the initial conditions $Y_\chi\left(T_{\rm RH}^{\rm max}\right) = Y_\phi\left(T_{\rm RH}^{\rm max}\right) = 0$. We recall that $M_1 = 5.10 \times 10^{10}~\mathrm{GeV}$, $M_2 = 3.28 \times 10^{14}~\mathrm{GeV}$ and $T_{\rm RH}^{\rm max} \simeq  6.5 \times 10^{15}~\mathrm{GeV}$.}
\end{center}
\end{figure}
\begin{figure}[t!]
\begin{center}
\subfigure[]{\includegraphics[width=0.48\textwidth]{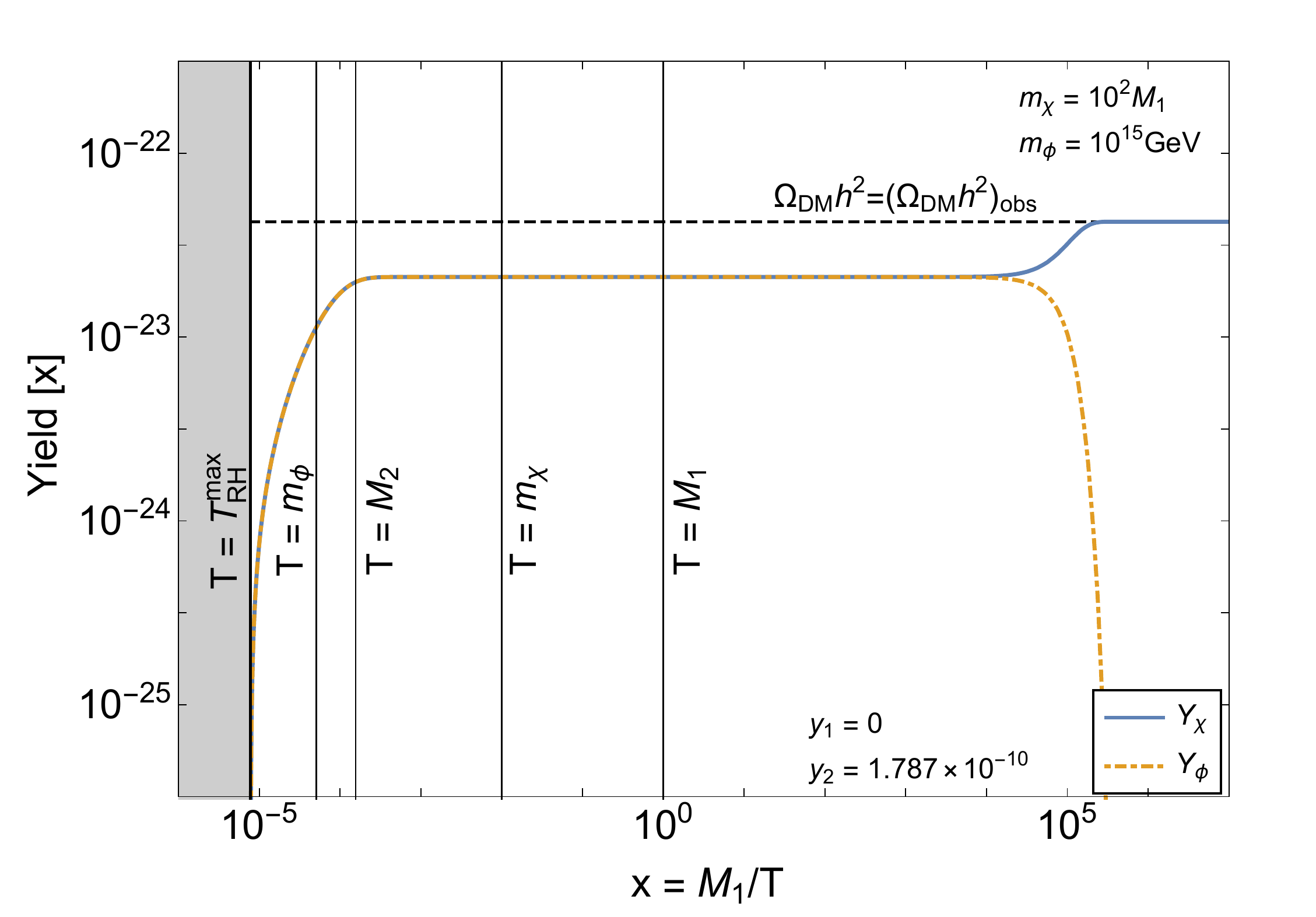}}
\subfigure[]{\includegraphics[width=0.48\textwidth]{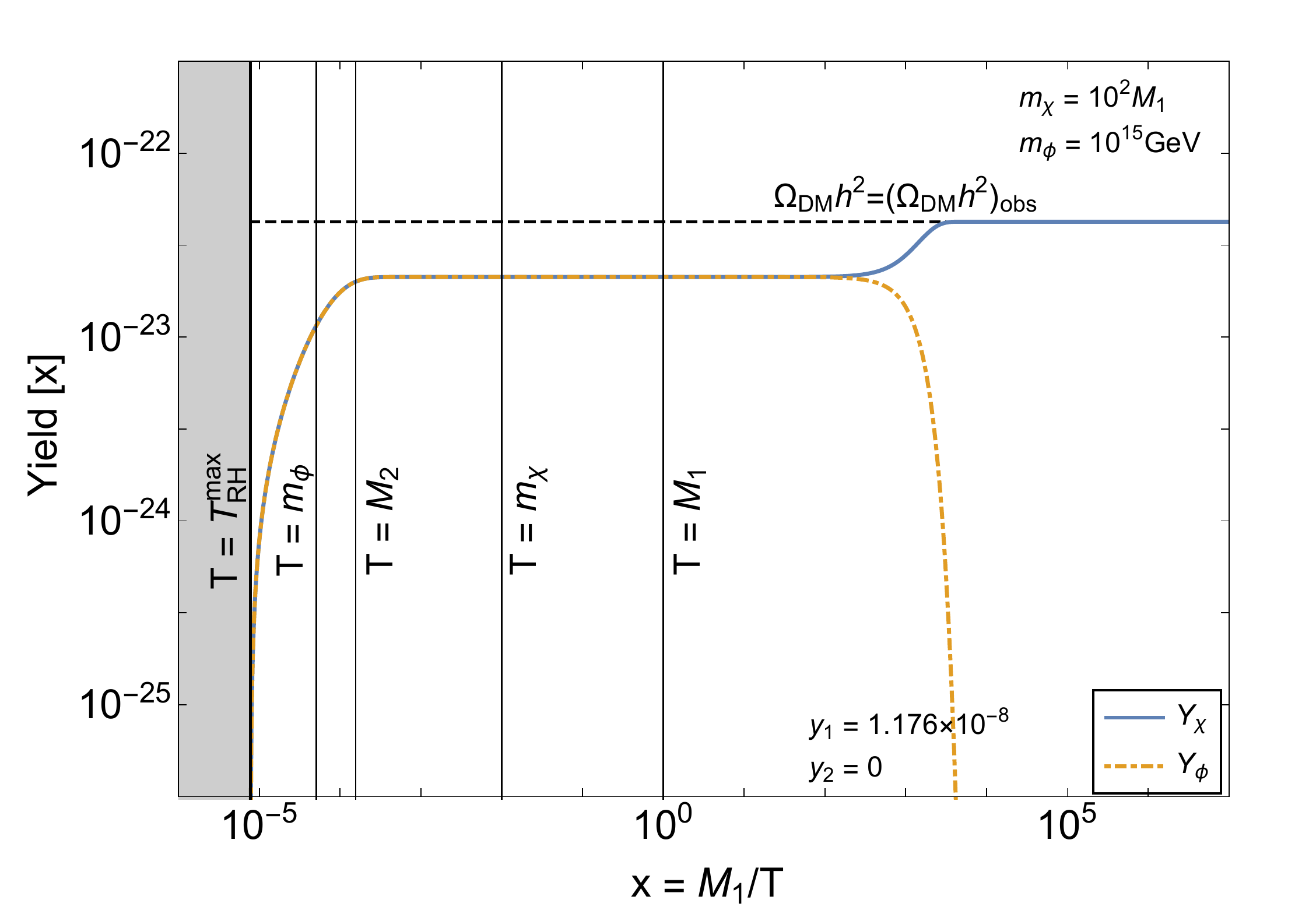}}
\subfigure[]{\includegraphics[width=0.48\textwidth]{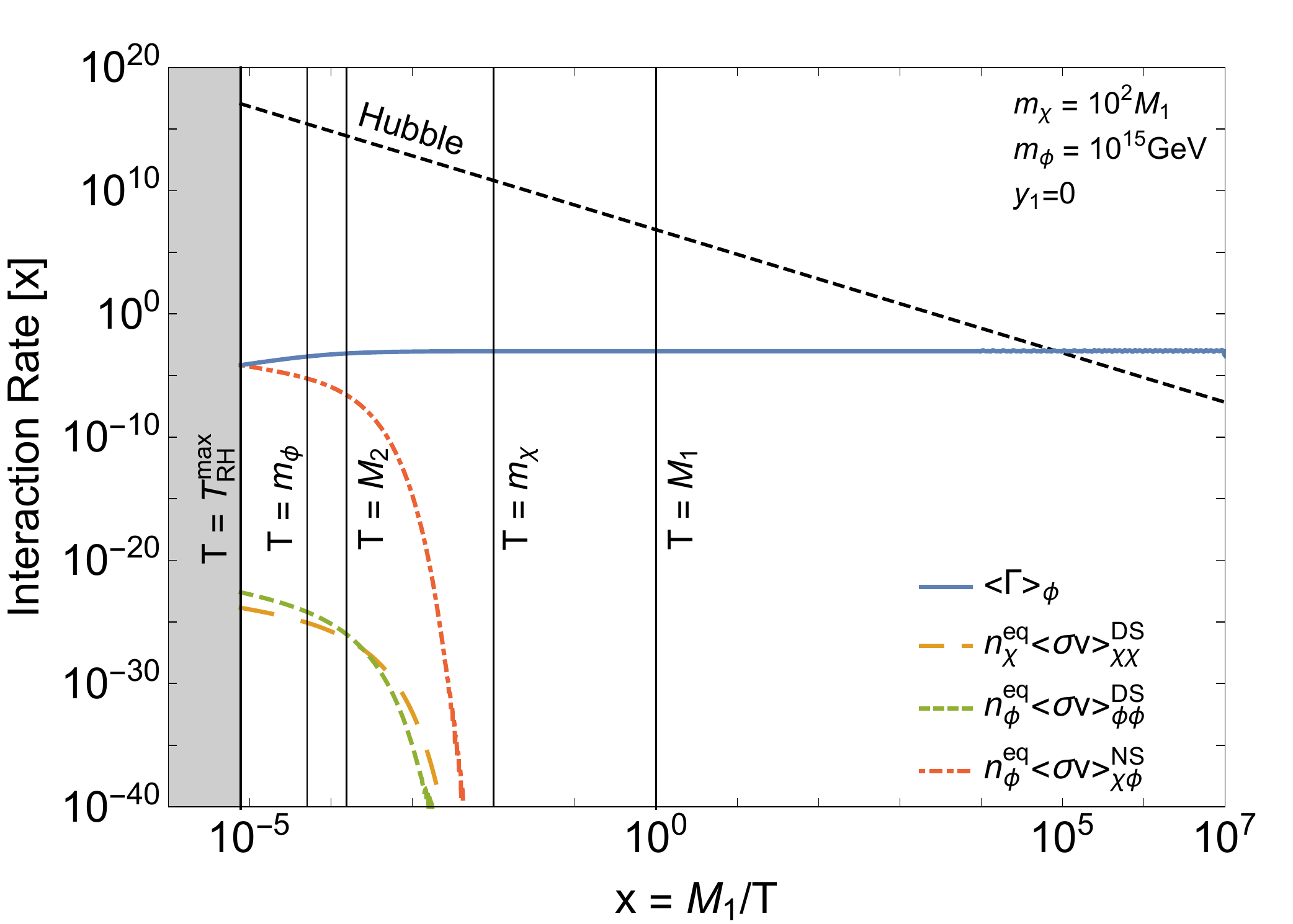}}
\subfigure[]{\includegraphics[width=0.48\textwidth]{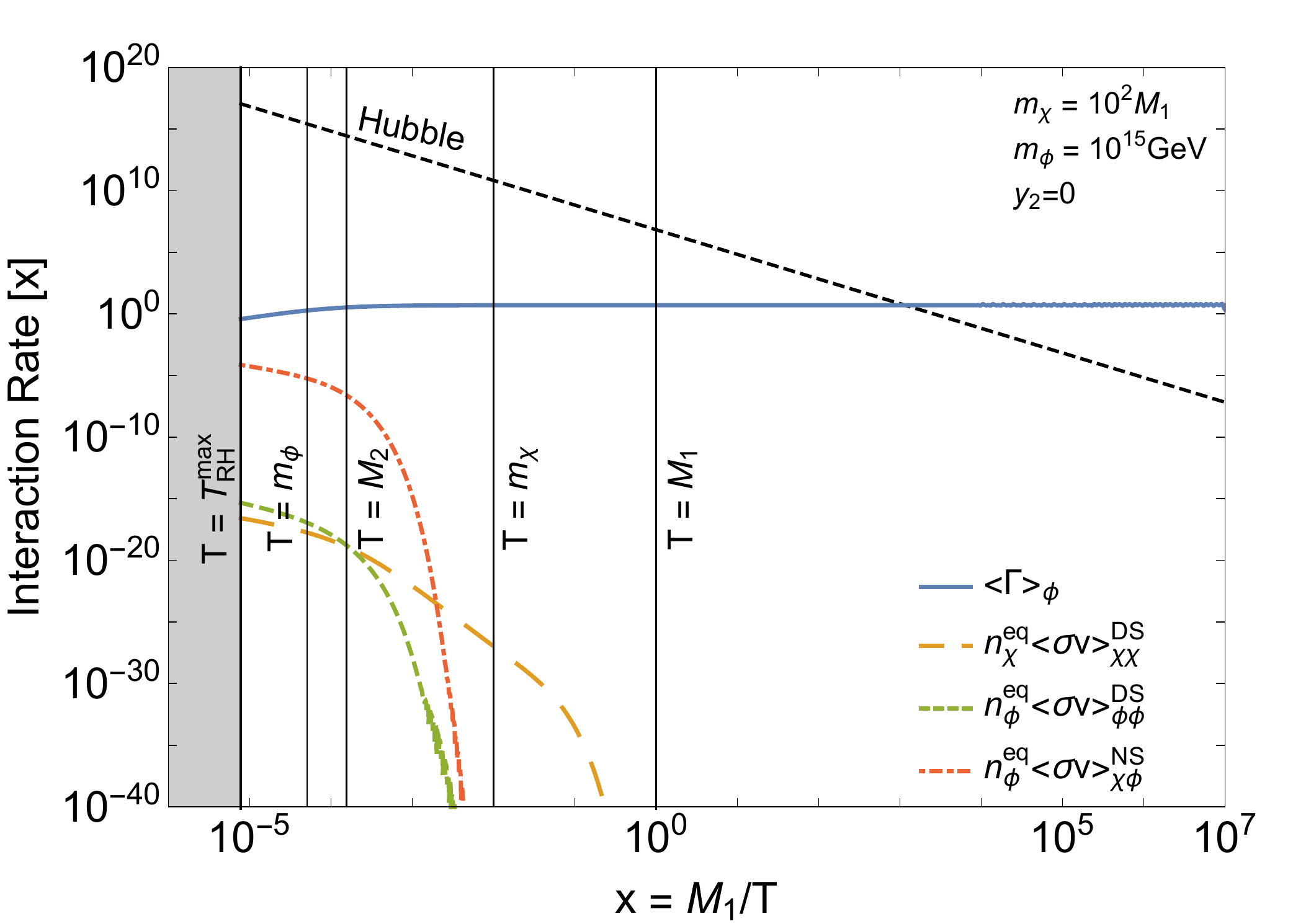}}
\caption{\label{fig:IntRate}The dark sector yields (upper panels) and the corresponding interaction rates (lower panels) of the main processes responsible for DM production as a function of $x = M_1 / T$. These plots correspond to the specific benchmark scenario of $m_\chi = 10^2 M_1$ and $m_\phi = 10^{15}~{\rm GeV}$. The left (right) panels correspond to $y_{1}=0$ ($y_{2}=0$). We recall that $M_1 = 5.10 \times 10^{10}~\mathrm{GeV}$, $M_2 = 3.28 \times 10^{14}~\mathrm{GeV}$ and $T_{\rm RH}^{\rm max} \simeq  6.5 \times 10^{15}~\mathrm{GeV}$.} 
\end{center}
\end{figure}

In Fig.~\ref{fig:Yield} we show the yields of DS particles obtained by solving the Boltzmann equations for two different choices of the masses $m_\chi$ and $m_\phi$. The DS coupling $y_1$ is set to be zero, which means the other coupling $y_2$ reaches its maximum value. By comparing the plots, it can be figured out that the decay of $\phi$ occurs after the freeze-in of the dark sector. At this stage, the yield of $\phi$ particles decreases while the one of $\chi$ particles increases, so matching the correct DM relic abundance highlighted by the horizontal dashed black line. In all these benchmark scenarios, the qualitative behavior of DS yields is unchanged. This is mainly due to the fact that a zero initial abundance for $\chi$ and $\phi$ particles is set at $T_{\rm RH} = T^{\rm max}_{\rm RH}$.

In Fig.~\ref{fig:IntRate} we focus on the benchmark scenario given by $m_\chi = 10^2 M_1$ and $m_\phi = 10^{15}~{\rm GeV} \simeq 2\times 10^4 M_1$. The upper panels display the yield as a function of the auxiliary variable $x=M_1 / T$ while in the lower panels we report the interaction rates of all the main processes entering the Boltzmann equations. As can be clearly seen, all the interaction rates are smaller than the Hubble parameter (dashed black lines) so justifying the assumption of the freeze-in regime. In these plots, the solid blue lines refer to the decay of scalar particles: the two-body decays of $\phi$ particles occur at the temperature for which the blue lines cross the Hubble parameter. This means that the two-body decays become efficient as compared to the expansion rate of the Universe, which is set by $\mathcal{H}$. In the figure, the left and right plots correspond to the case where $y_1 = 0$ and $y_2 = 0$, respectively. Remarkably, in all these benchmark scenarios, the DM production is driven by the neutrino Yukawa interactions as can be seen by the fact that the NS interaction rates (red dot-dashed lines) are larger than the DS ones (green dashed and yellow long-dashed lines). This implies that in this model, there exists an intriguing relation between the neutrino and the dark sectors. As will be discussed in the next subsection, this relation is no longer achieved for low reheating temperatures.

\subsection{ Reheating temperature $T_{RH}$\label{sec:TRH}}
\begin{figure}[t!]
\begin{center}
\subfigure[]{\includegraphics[width=0.47\textwidth]{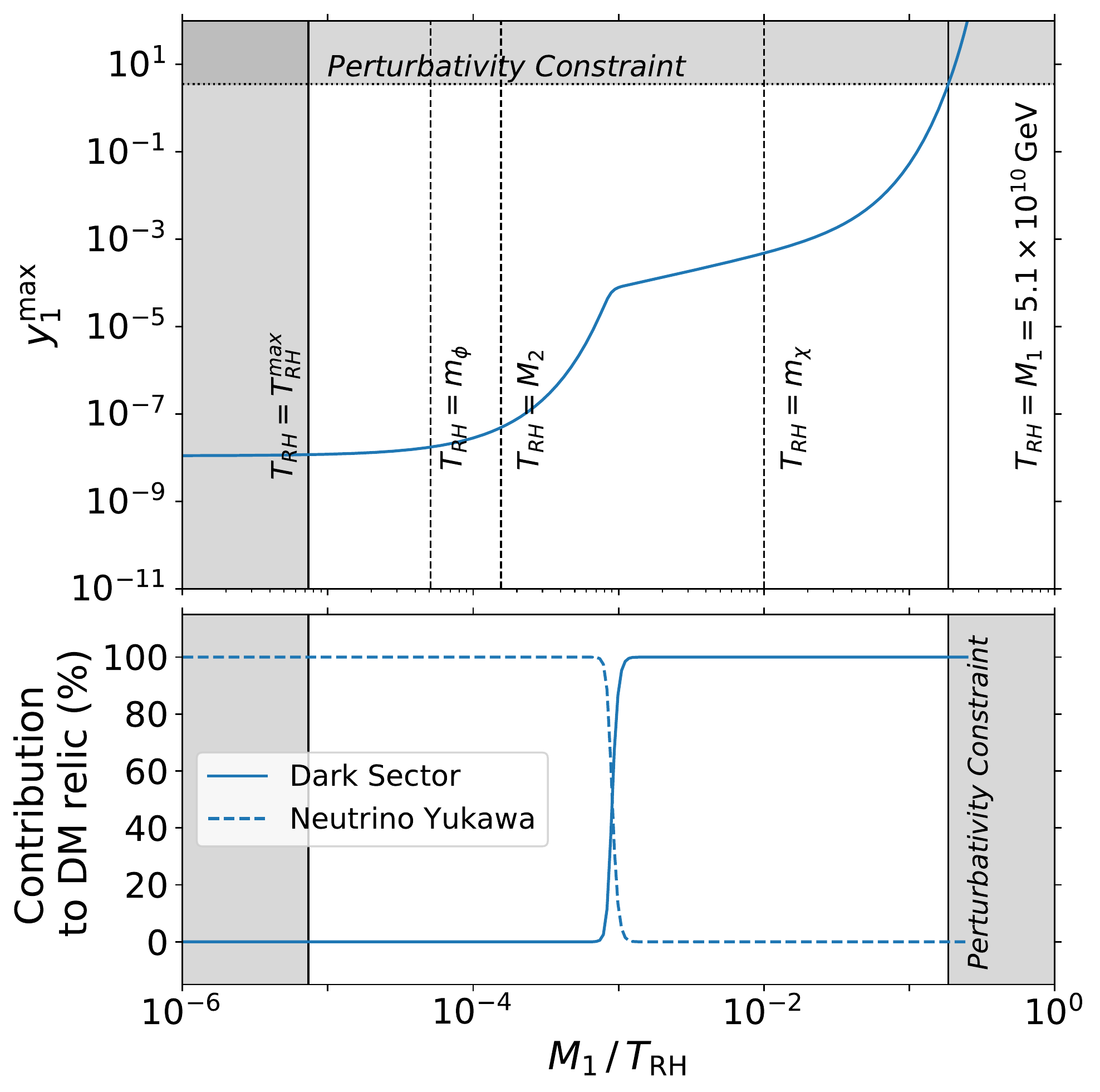}}
\hskip5.mm
\subfigure[]{\includegraphics[width=0.47\textwidth]{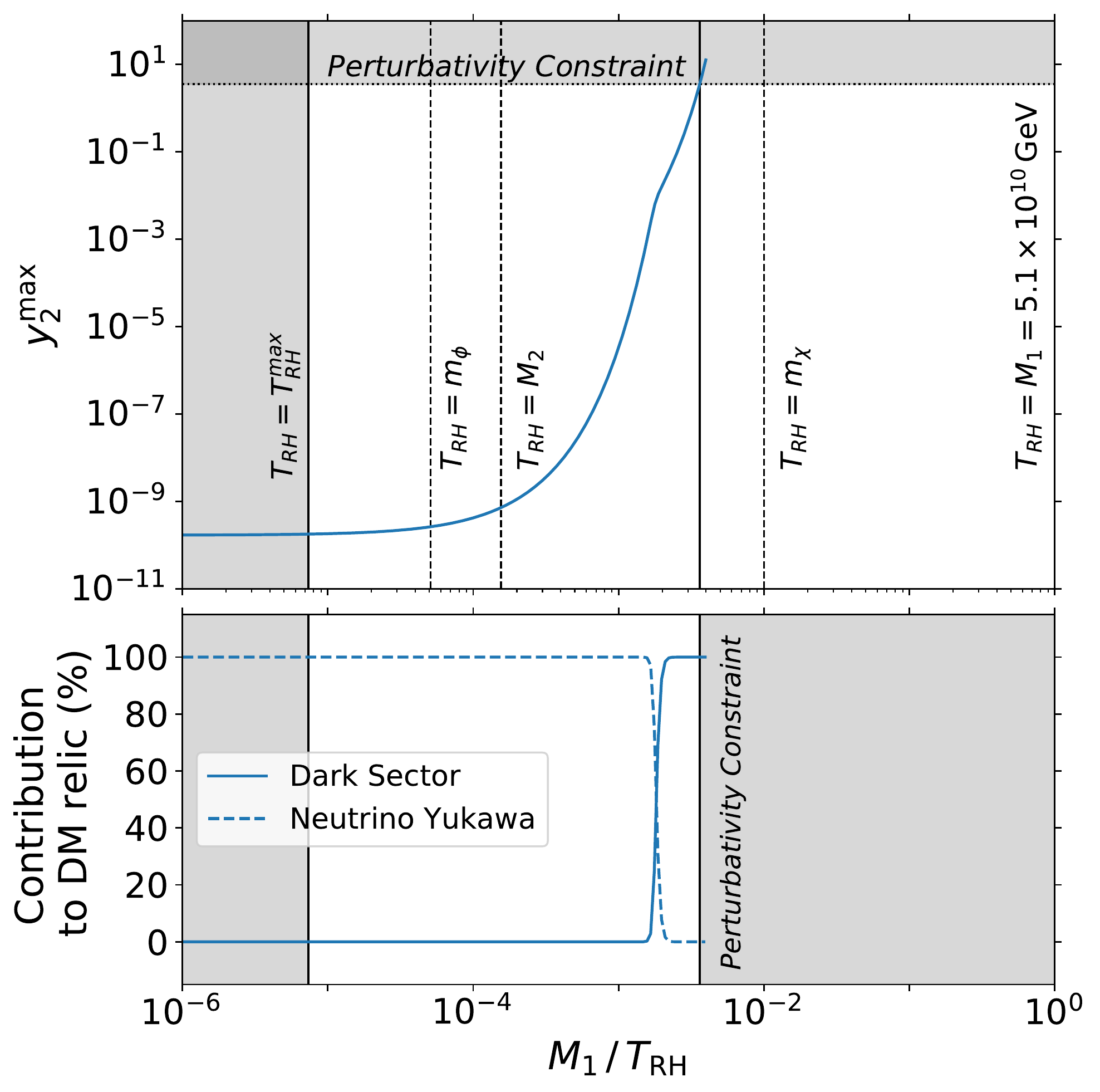}}
\caption{Maximum values for the RHN portal
couplings, $y_1^{\rm max}$ (left panel) and $y_2^{\rm max}$ (right panel), as a function of the reheating temperature $T_{\rm RH}$, for the benchmark case of $m_\chi=10^2 M_1$ and $m_\phi = 10^{15}~{\rm GeV}$. The gray shaded areas display the excluded regions from perturbativity and the requirement $T_{\rm RH} \leq T_{\rm RH} ^{\rm max}$. The lower part of the plots shows the relative contribution of DS and NS scatterings to the DM relic abundance.
\label{fig:yTi}}
\end{center}
\end{figure}

In Fig.~\ref{fig:yTi} we report how the upper limits of dark sector couplings depend on the reheating temperature for the benchmark scenario of $m_\chi = 10^2 M_1$ and $m_\phi = 10^{15}~{\rm GeV}$. It can be observed that the maximum values of the DS couplings remain constant when the reheating temperature is large enough. The limits of couplings start to increase when the reheating temperature becomes smaller than the mass of the scalar dark particle because of the kinematical suppression on neutrino Yukawa interactions. At this stage, the neutrino Yukawa interactions dominate the dark matter production (as highlighted in the lower part of the plots). Then there is a critical point in each panel, where the increasing tendency of the coupling suddenly changes. These critical points correspond to the equality between the contributions to the DM relic from the neutrino Yukawa interactions and the dark sector interactions (more specific, the $\chi\chi$ scattering). At higher reheating temperatures, the NS interactions dominate the DM production, while at lower reheating temperatures the production is driven only by the dark sector. After the critical point, $y_1^{\rm max}$ experiences a more gentle and stable increase until the $\chi\chi$ scattering is kinematically suppressed, while $y_2^{\rm max}$ only slows down its increase slightly. Finally, at very low reheating temperatures the DS couplings required to achieve the DM relic abundance are larger the perturbativity constraint defined by $y \leq \sqrt{4\pi}$. Such an intrinsic upper bound is displayed by horizontal dashed black lines.

\begin{figure}[t!]
\begin{center}
\subfigure[]{\includegraphics[width=0.48\textwidth]{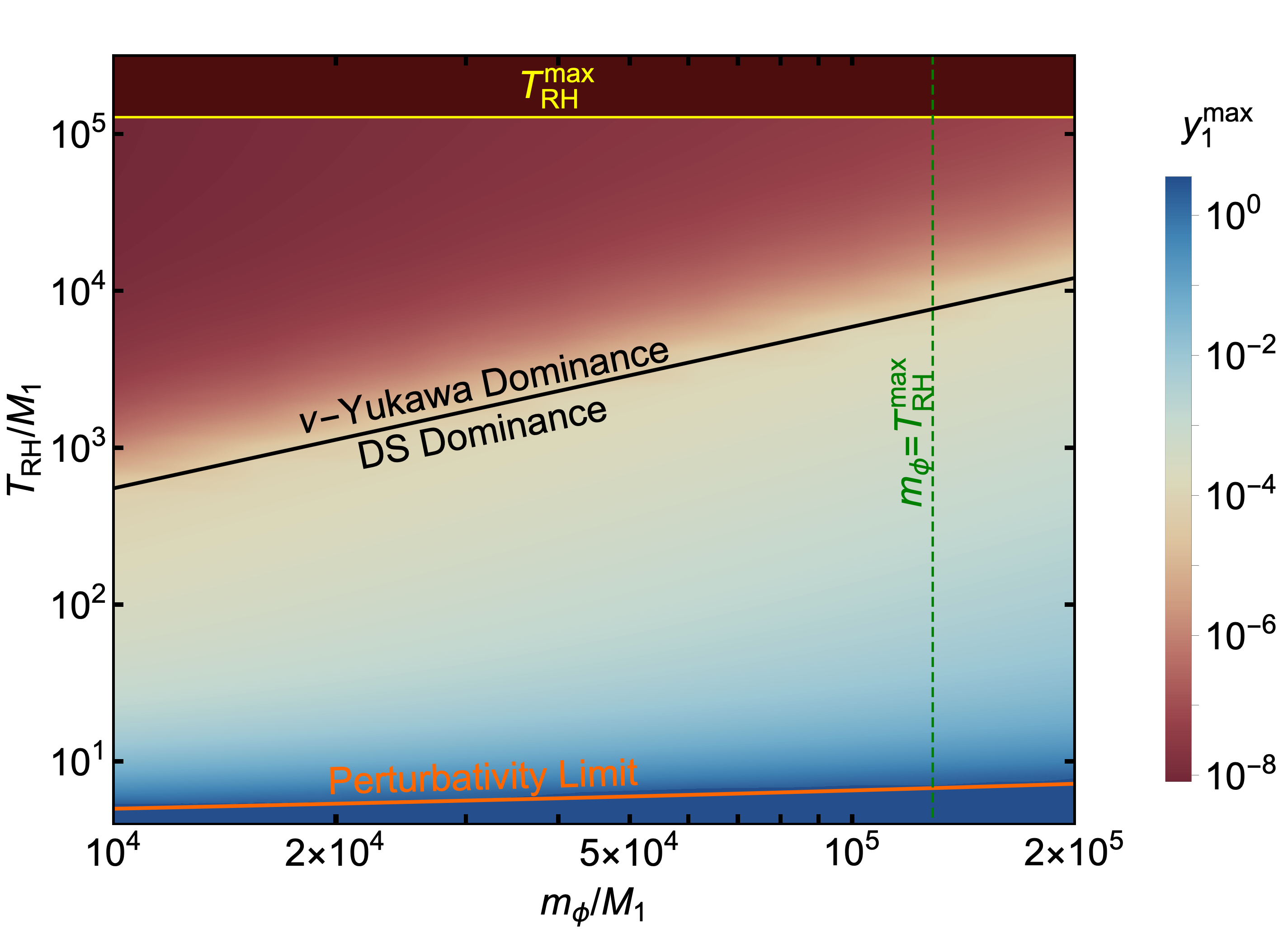}}
\subfigure[]{\includegraphics[width=0.48\textwidth]{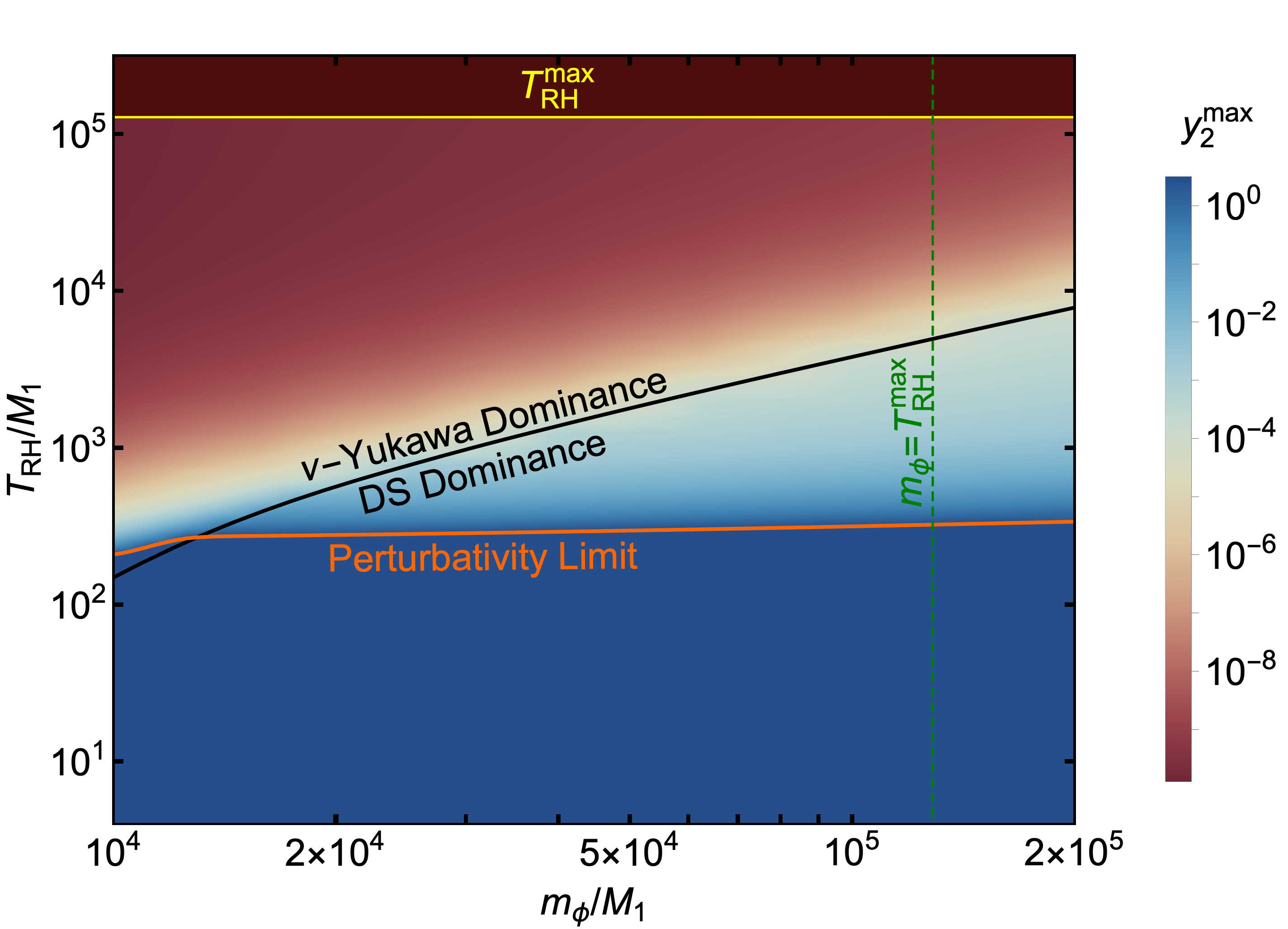}}
\subfigure[]{\includegraphics[width=0.48\textwidth]{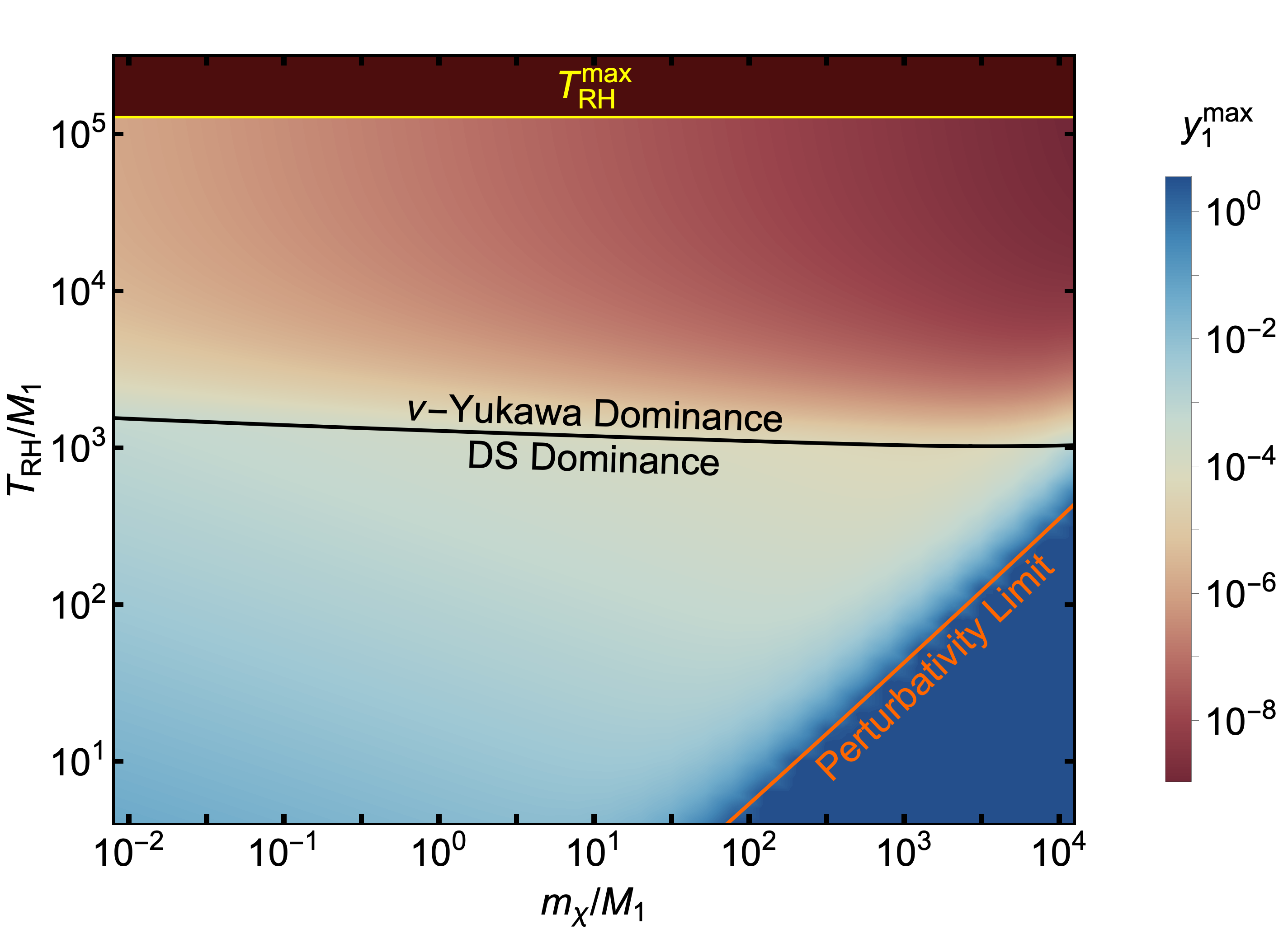}}
\subfigure[]{\includegraphics[width=0.48\textwidth]{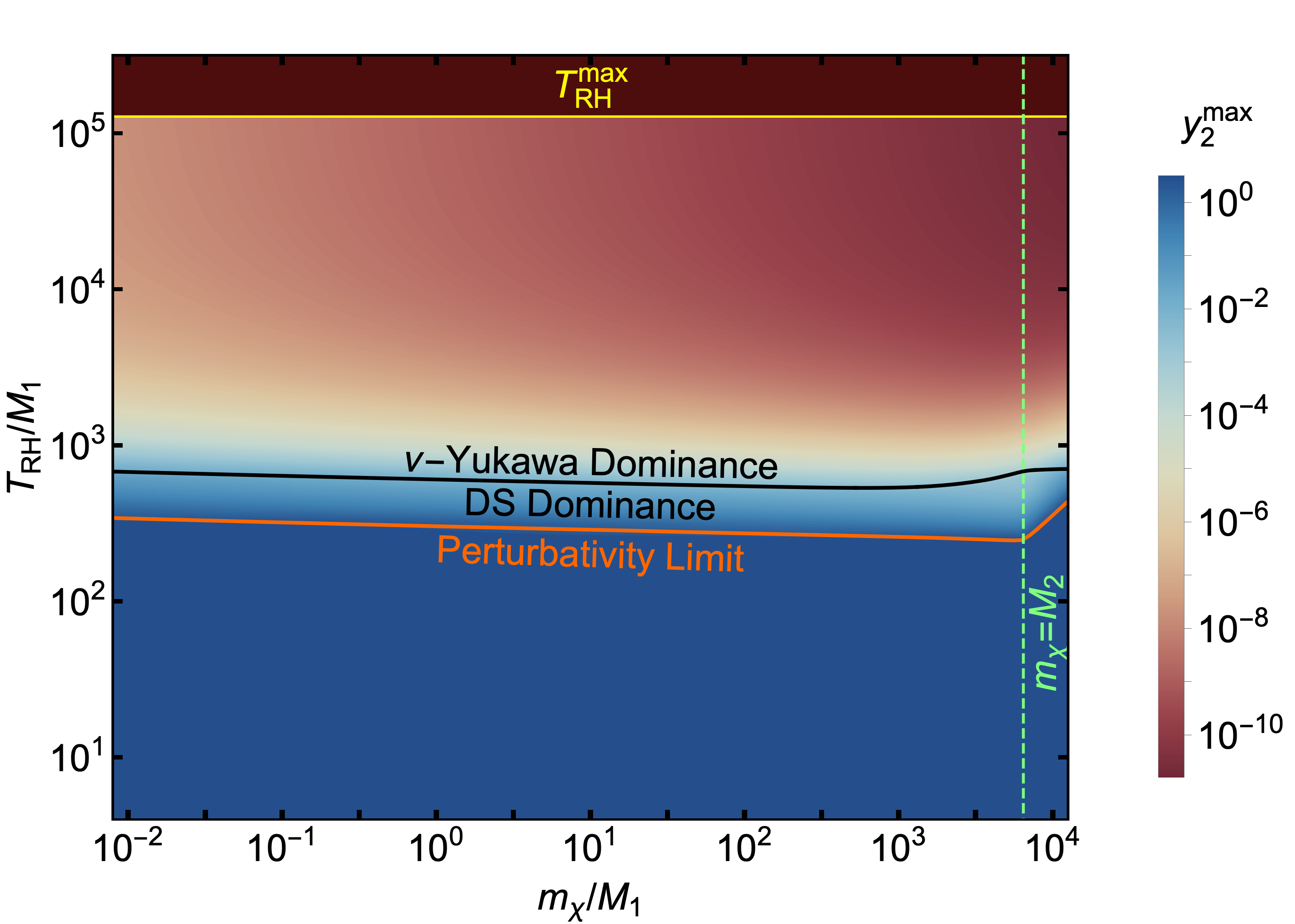}}
\caption{Maximum values for the RHN portal couplings as a function of dark sector masses (in units of $M_1$) and reheating temperature (in units of $M_1$). In the upper (lower) plots we set $m_\chi = 10^2 M_1$ ($m_\phi = 10^{15}$GeV). The left and right panels show the maximum allowed value for the couplings $y_1$ and $y_2$, respectively. The parameter space is divided as follows: the region above yellow line is excluded by the requirement $T_{\rm RH} \leq T_{\rm RH}^{\rm max} \simeq  6.5 \times 10^{15}~{\rm GeV}$; the region below the orange line is excluded by the perturbativity of the couplings; the region above and below the black line represents respectively the dominance of neutrino Yukawa scatterings and dark sector process in driving dark matter production. We recall that $M_1 = 5.10 \times 10^{10}~\mathrm{GeV}$ and $M_2 = 3.28 \times 10^{14}~\mathrm{GeV}$.
\label{fig:yTmPhi}}
\end{center}
\end{figure}

In Fig.~\ref{fig:yTmPhi} we report the main results of the present paper obtained by a full scan of the model parameter space by fixing $m_\chi = 10^2 M_1$ (upper plots) and $m_\phi = 10^{15}~{\rm GeV}$ (lower plots).  The plots show the maximum allowed values for the DS couplings, $y_1^{\rm max}$ (left plots) and $y_2^{\rm max}$ (right plots), for different values of the dark particles masses and the reheating temperature. The bluer the color, the larger the DS coupling. The dark blue region, bounded from above by the solid orange line, highlights the region where the perturbation theory breaks. We have checked that the DM production always proceeds through the freeze-in mechanism, even though the couplings are very large for low reheating temperature. The interactions rates are indeed suppressed either by the smallness of the coupling or by approaching kinematic energy thresholds. Hence, they are smaller than the Hubble parameter, implying that the dark sector is never in thermal equilibrium with the thermal bath. Eventually, it is worth noticing that all the plots are divided into two regions by a solid black line. This curve marks the parameters for which there is the equality between the contributions of the DS and NS scatterings to the DM relic abundance. Above the black lines, the neutrino Yukawa scatterings drive the DM production and the relation between dark matter and neutrinos is preserved. For low reheating temperature (large DS couplings), the DS scatterings dominate so spoiling the DM-$\nu$ relation.

\section{Conclusions\label{sec:con}}

In this paper, we have proposed a minimal model that at the same time provides a realistic neutrino spectrum through a type-I seesaw mechanism, the correct baryon asymmetry of the Universe via leptogensis, and a viable non-WIMP dark matter candidate. In particular, we have studied a minimal extension of the Littlest Seesaw model with two dark particles charged under a global $U(1)_{D}$. The dark sector and the seesaw sector are connected through the right-handed neutrino (RHN) portal. The novelties of the present analysis with respect to the one presented in Ref.~\cite{Chianese:2018dsz} are threefold:
\begin{itemize}
\item We have considered the additional constraint of explaining the baryon asymmetry of the universe through leptogenesis. As studied in Ref.~\cite{King:2018fqh}, such a requirement, along with fitting the neutrino mass and mixing parameters and taking into account renormalization group equations, completely fixes the neutrino Yukawa sectors. In particular, the benchmark Littlest seesaw model considered here requires the masses of the two right-handed neutrinos to be different, i.e. $M_1=5.10 \times 10^{10}~\rm GeV$ and $M_2=3.28 \times 10^{14}~\rm GeV$;
\item We allow more freedom in the model considering the two RHN portal couplings to be different. This naturally emerges from having non-degenerate neutrino masses;
\item We study the impact of the initial conditions of the Universe on the dark matter production, that is achieved through the freeze-in mechanism. In particular, we provide the allowed parameter space of the dark sector achieving the correct DM abundance for different values of the reheating temperature. In particular, we considered the range $M_1 \leq T_{\rm RH} \lesssim 6.5\times 10^{15}~{\rm GeV}$ where the lower and upper bounds are respectively defined by the requirement of successful leptogenesis and the Planck~\cite{Akrami:2018odb} and BICEP-Keck~\cite{Ade:2018gkx} measurements .
\end{itemize}
The parameters of the dark sector, namely the two RHN portal couplings and the two masses, are constrained by solving the Boltzmann equations and requiring that the dark Dirac fermion $\chi$ is the dark matter. Motivated by the heavy right-handed neutrino masses required for leptogenesis in the LS model, we have focused on very heavy FIMPzilla DM particles, with a mass of the same order or even larger than $10^{10}$~GeV. When the reheating temperature is fixed to be equal to its maximum, we have confirmed the existence of a relation between neutrino physics and dark matter via the neutrino Yukawa couplings. In particular, the DM production is driven by neutrino Yukawa interactions whose couplings are completely fixed by neutrino physics and leptogenesis, and which dominate over the  RHN portal couplings. In this framework, we have investigated the maximum allowed values for the RHN portal couplings required to achieve the correct DM relic density, as a function of the masses of dark particles. Going beyond the previous analysis~\cite{Chianese:2018dsz}, we have shown that the DM-$\nu$ relation strongly depends on the value of the reheating temperature of the Universe. In particular, the lower the reheating temperature, the larger the DS couplings required to correctly produce DM particles. In this case, therefore, the DS scatterings in Fig.~\ref{fig:Feyn1}(a) dominate the DM production, while the neutrino Yukawa ones in Fig.~\ref{fig:Feyn1}(b) are subdominant. However, such regions in the model parameter space might be interesting for phenomenological studies due to the very large RHN portal couplings.

\section*{Acknowledgments}
We thank Matteo Biagetti and Mathias Pierre for useful discussions and comments. SFK acknowledges the STFC Consolidated Grant ST/L000296/1 and the European Union's Horizon 2020 Research and Innovation programme under Marie Sk\l {}odowska-Curie grant agreements Elusives ITN No.\ 674896 and InvisiblesPlus RISE No.\ 690575. BF acknowledges the Chinese Scholarship Council (CSC) Grant No.\ 201809210011 under agreements [2018]3101 and [2019]536.

\appendix

\section{Amplitudes}
\label{A}

In this Appendix, we report the squared matrix elements for all the scattering processes that are important for the DM production as a function of the corresponding Mandelstam variables $s$, $t$ and $u$. In the following computations, we have used the Feynman rules for Majorana fermions reported in Ref.~\cite{Denner:1992me,Denner:1992vza}. For the DS processes drawn in Fig.~\ref{fig:Feyn1} we have
{\small
\begin{eqnarray}
\overline{\left|\mathcal{M}\right|^2}_{\phi\phi^*\rightarrow n_in_j} & = & - \frac{\left|y_{i}\, y_{{\rm DS}j}\right|^2}{\left(t-m_\chi^2\right)^2} \left[\left(t-m_\phi^2+M_{Ri}^2\right)\left(t-m_\phi^2+M_{Rj}^2\right)+t\left(s-M_{Ri}^2-M_{Rj}^2\right)\right] \nonumber \\
& & - \frac{\left|y_{{\rm DS}i}\, y_{j}\right|^2}{\left(u-m_\chi^2\right)^2} \left[\left(u-m_\phi^2+M_{Ri}^2\right)\left(u-m_\phi^2+M_{Rj}^2\right)+u\left(s-M_{Ri}^2-M_{Rj}^2\right)\right]  \\
& & \mp \frac{2 M_{Ri} M_{Rj} \,{\rm Re}\left(y_{i}\, y_{j}^*\right)}{\left(t-m_\chi^2\right)\left(u-m_\chi^2\right)} \left(s+t+u-2M_{Ri}^2-2M_{Rj}^2 \right) \nonumber \,, \\
&& \nonumber \\
\overline{\left|\mathcal{M}\right|^2}_{\chi\overline{\chi}\rightarrow n_in_j} & = &\left|y_{i}\, y_{j}\right|^2 \left[\frac{\left(t-m_\chi^2-M_{Ri}^2\right)\left(t-m_\chi^2-M_{Rj}^2\right)}{4\left(t-m_\phi^2\right)^2} +\frac{\left(u-m_\chi^2-M_{Ri}^2\right)\left(u-m_\chi^2-M_{Rj}^2\right)}{4\left(u-m_\phi^2\right)^2} \right] \nonumber \\
& & \mp \frac{M_{Ri} M_{Rj} \,{\rm Re}\left(y_{i}\, y_{j}^*\right)\,\left(s-2m_\chi^2\right)}{4\left(t-m_\phi^2\right)\left(u-m_\phi^2\right)} \,,
\end{eqnarray}}
where the sign $-$ ($+$) in the interference term is in case of equal (different) right-handed neutrinos, i.e. $i=j$ ($i \neq j$). On the other hand, neglecting the neutrino and lepton masses as well as the Higgs mass, the sum of the squared matrix elements for all the $\nu-$Yukawa scattering processes take the expression
{\small
\begin{eqnarray}
\overline{\left|\mathcal{M}\right|^2}_{\phi\overline{\chi}\rightarrow h^0\nu_i/G^0\nu_i/G^\pm l_i^\mp} & = & 
\frac{\left|y_{1}\right|^2 \tilde{y}_{\nu,1}^2}{\left(s-M_{R1}^2\right)^2} 
\left[\left(s+m_\chi^2-m_\phi^2\right)s+
\left(s-M_{R1}^2\right)\left(t-m_\chi^2\right)\right] 
\nonumber \\
& & +  
\frac{\left|y_{2}\right|^2 \tilde{y}_{\nu,2}^2}{\left(s-M_{R2}^2\right)^2}
\left[\left(s+m_\chi^2-m_\phi^2\right)s+
\left(s-M_{R2}^2\right)\left(t-m_\chi^2\right)\right] \\
& & + 
\frac{2\,{\rm Re}\left(y_{1} \, y_{2}^*\right) \tilde{y}_{\nu, 12}^2}{\left(s-M_{R1}^2\right)\left(s-M_{R2}^2\right)}
\left[\left(s+m_\chi^2-m_\phi^2\right)s+ 
\left(s-M_{R1}M_{R2}\right)\left(t-m_\chi^2\right)\right] \nonumber \,, \\
\nonumber
\end{eqnarray}
where the  analytical expressions of the effective squared Yukawa couplings are given by
\begin{eqnarray}
\tilde{y}_{\nu,1}^2 &=& \sum_{i=1}^3 \left(2\left|\left(U_\nu^\dagger Y\right)_{i1}\right|^2+\left|Y_{i1}\right|^2\right) \nonumber\\
\tilde{y}_{\nu,2}^2 &=& \sum_{i=1}^3 \left(2\left|\left(U_\nu^\dagger Y\right)_{i2}\right|^2+\left|Y_{i2}\right|^2\right) \nonumber\\
\tilde{y}_{\nu,12}^2 &=& 2\sum_{i=1}^3 \left[2{\rm Re}\left(\left(U_\nu^\dagger Y\right)_{i1} \left(U_\nu^\dagger Y\right)^{*}_{i2}\right)+{\rm Re}\left(Y_{i1} Y^{*}_{i2}\right)\right]\nonumber
\end{eqnarray}

\bibliography{DarkMatter}

\end{document}